\documentclass[12pt]{book}
\usepackage{sussp} 
\usepackage{graphicx}     
\usepackage{makeidx}
\makeindex
\graphicspath{{./figs/}}

\newcommand{\nub}{\bar{\nu}}
\newcommand{\qbar}{\bar{q}}
\newcommand{\rmt}{\rm\textstyle}

\begin{document}

\headings{Neutrino Interactions} 
{Neutrino Interactions}
{Kevin McFarland}
{University of Rochester, Rochester, NY, USA 14627} 





\section{Introduction and Motivations}

The study of neutrino interaction physics played an important role in
establishing the validity of the theory of weak interactions and
electroweak unification.  Today, however, the study of interactions of
neutrinos takes a secondary role to studies of the properties of
neutrinos, such as masses and mixings.  This brief introduction
describes the historical role that the understanding of neutrino
interactions has played in neutrino physics and what we need to
understand about neutrino interactions to proceed in future
experiments aimed at learning more about neutrinos.

The original application of neutrino interactions was the discovery of
the neutrino itself.  For most physicists today, who came of age
professionally well after the first observation of neutrinos, it takes a bit
of thought to understand the perspective of the experimenters seeking
to discover the neutrino.  A close analogy today might be the search
for interactions of weakly interacting massive (WIMP) dark matter
particles.  In order to sensibly design an experiment to search for a
new particle and to interpret the results, an experimenter needs
guidance about the probable type and rate of interactions to be
observed.  For the case of WIMP dark matter, information about the
strength of interactions comes from the standard cosmological model
which relates modern day abundance of dark matter to production and
annihilation cross sections.
\index{dark matter}

\index{Fermi theory of weak interactions}
In the case of neutrinos in the early 1950s, the guiding principle was
the Fermi ``four fermion'' theory of the weak interaction (Fermi
1934).  This theory introduced a four-fermion vertex connecting
a neutron $n$, a proton $p$, an electron $e^-$ and an anti-neutrino, 
$\nu$ or $\bar{\nu}$ to explain
neutron decay, $n\to p e^-\bar{\nu}$ in terms of a single unknown
coupling constant, $G_F$.  Because that single constant governed the
strength of all weak interactions among these particles, the Fermi
theory led to definite prediction for neutrino interactions involving
these particles. The prediction for the cross section of $\bar{\nu} p \to e^+ n$ was
first derived by Bethe and Peierls shortly after the Fermi theory was
published (Bethe and Peierls 1934).  
For neutrinos with energies of a few MeV from a
reactor, a typical cross section in this theory was predicted to be
$\sigma_{\bar{\nu} p}\sim 5\times10^{-44}$~cm$^2$.  Interestingly,
this prediction for reactor neutrino cross sections is still accurate
today, up to a factor of two required to account for the then
unknown phenomenon of maximal parity violation in the weak
interaction!
This small cross section is, as we all recognize today,
the primary challenge in performing experiments with neutrinos.  By
contrast, the cross section for the 
corresponding electromagnetic process with a photon $\gamma$ 
at similar energies 
is $\sigma_{\gamma p}\sim 10^{-25}$~cm$^2$. 
The tiny neutrino cross section means that the
mean free path of reactor neutrinos with energies of a few MeV in steel is
approximately ten light years.

With these predictions in place, the stage was set for the two
critical measurements establishing the existence and nature of the
neutrinos from nuclear reactors: the Davis {\em et al} null
measurement of the reaction $\bar{\nu}~+~^{37}Cl~\to~^{37}Ar~+~X$
and Reines and Cowan's observation of $\bar{\nu} p \to e^+ n$ in
1955-56.  In modern language, the latter measurement establishes the
existence of the neutrino and validates the universality of the Fermi
theory and the former non-measurement shows that the neutrino and
anti-neutrino carry an opposite conserved lepton number which forbids
$\bar{\nu} n \to e^- p$ (Reines 1996).


\subsection{A Cautionary Tale: Discovery of the Weak Neutrino Current}

\begin{figure}
\centering
\includegraphics[width=6cm]{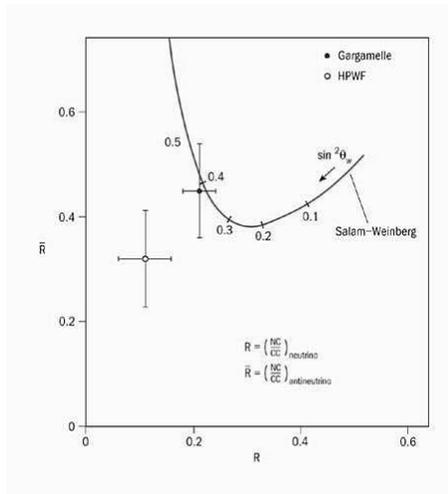}
\caption{Measurement of $R^{\bar{\nu}}$ vs.\ $R^{\nu}$ from the
  Gargamelle and HWPF collaborations compared with the prediction of the
  electroweak standard model.}
\label{fig:weinberg-rnunubar}
\end{figure}

A more sobering story involving knowledge of neutrino cross sections
involves the discovery of the weak neutral current in neutrino
interactions.  No textbook would be complete without the requisite
picture of the famous single electron event in the Gargamelle bubble
chamber, attributed to $\bar{\nu}_e e^- \to \bar{\nu}_e e^-$.  While
this event is a wonderful illustration of a weak neutrino process, it
was not the discovery channel.  As we will see, the cross section for
this reaction is exceedingly small, and concerns about backgrounds and
the lack of corroborating information in such a reaction make it a
difficult channel in which to claim a discovery.  The discovery
measurement for the weak neutral current involves processes where
neutrinos scatter off of the nuclei in the target allowing the
experimenters to measure a quantity such as 
\begin{equation}
R^{\nu}=\frac{\sigma(\nu_\mu N \to \nu_\mu X)}{\sigma(\nu_\mu N \to \mu^- X)} 
\end{equation}
or its analog with an anti-neutrino beam.  Figure~\ref{fig:weinberg-rnunubar}
shows these two measurements compared with the prediction of the
electroweak standard model as a function of its single parameter
not constrained by low energy data,
$\sin^2\theta_W$, which is the weak mixing angle or Weinberg angle.
\index{neutral current interaction}

This major triumph for the standard model of electroweak unification
was sadly complicated by an involved saga which ultimately boiled down
to uncertainties in translating observed events to the measurement of
$R^\nu$.  Experimentally, the measurement consists of identifying
events as either containing or not containing of final state muon and
using this distinguishing feature to separate charged and neutral
current interactions.  Very low energy muons are difficult to separate
from other particles, primarily charged mesons, produced in inelastic
scattering from nuclei, and so these events constitute a background to
the neutral current sample.  Equally problematic for this measurement
are neutral current events which produce charged hadrons in the final
state that are confused with energetic muons.  Without a good model
for the production of these charged mesons or a good understanding of
the probability of confusing charged mesons with muons in the
detector, the experimental problem of isolating sufficiently clean
samples with high statistics hobbled efforts to produce a convincing
observations by both of the competing collaborations, Gargamelle at
CERN and HWPF at Fermilab (Galison 1983).  It is notable that this
important discovery was never honored with a Nobel prize, despite its
critical role in validating the electroweak theory.


\subsection{Cross Section Knowledge and Next Generation Oscillation
  Experiments}

\index{neutrino oscillations}
The current and next generation of accelerator neutrino oscillation experiments
are again facing limitations arising from knowledge of neutrino
cross sections.  The physics roadmap of precisely measuring the
``atmospheric'' oscillation parameters, measuring $\theta_{13}$,
determining the neutrino mass hierarchy
and measuring the CP violating phase, $\delta$, has driven an experimental
program to be realized in several steps.  Currently this program
is the measurement of $\nu_\mu \to \nu_\mu$ transition probabilities in
wide band beams with baseline $L$ and mean energies $E$ near $L/E\sim400$~km/GeV (K2K and
MINOS), and the measurement of $\nu_\mu \to \nu_\tau$ near $\tau$
production threshold (OPERA). In the near future, it includes
narrowband (off-axis) beam experiments again near $L/E\sim400$~km/GeV
to precisely measure $\nu_\mu \to \nu_e$ transitions in neutrino and
anti-neutrino beams (T2K and NOvA).  Most likely, completion of this
program will require a new generation of experiments to study these
transitions at the second oscillation maximum as well,
$L/E\sim1200$~km/GeV, either in narrow band beams (T2KK) or wideband
beams (discussed in FNAL to DUSEL proposals).  Practical
considerations limit the range of possible baselines to
$L\stackrel{<}{\sim}2000$~km because of available sites and achievable
event rates and $E\stackrel{>}{\sim}0.5$~GeV because of the roughly quadratic
drop in the signal cross section and because of significant nuclear
effects with neutrinos energies below this limit.  This
implies that the neutrinos to be studied will have $0.5<E_\nu<5$~GeV.
As we will see, this region is at the threshold for inelastic
interactions on nucleons, which is a particularly difficult energy
region to model and is lacking in data to contribute to understanding the
relevant effects governing the details of cross sections. 

\begin{figure}
\centering
\includegraphics[width=12cm]{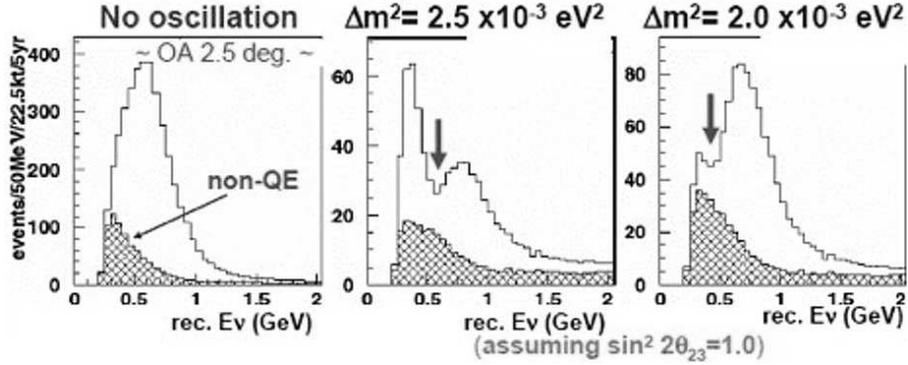}
\caption{The expected far detector $\nu_\mu$ candidate
  spectrum in the T2K experiment for $\theta_{23}=\pi/4$.  The hatched
  area in each plot shows expected backgrounds.}
\label{fig:t2k-disapp}
\end{figure}

\index{muon neutrino disappearance}
Knowledge of cross sections impacts a $\nu_\mu \to \nu_\mu$
disappearance measurement in this energy regime because, regardless of
experimental techniques, the details of the final state will impact
the separation of signal from background and the measurement of
neutrino energy in a given event.  Figure~\ref{fig:t2k-disapp}
illustrates the effect of backgrounds on the measurement of the
maximum oscillation probability on the T2K experiment.  If the
background to the signal, in this case primarily from inelastic
charged-current events, cannot be accurately estimated, then it
becomes difficult to measure the depth of the oscillation ``dip''
which is used to measure $\theta_{23}$.  In a broadband beam like that
of the MINOS experiment where the neutrinos at the energy of maximal
oscillation have an energy near $2$~GeV, the differences in energy
response between baryons, charged pions and neutral pions in the final
state lead to a significant uncertainty in reconstructed energy.  This
uncertainty in turn impacts the measurement of the energy of the
oscillation ``dip'' which determines $\delta m^2_{23}$.


\begin{figure}
\centering
\mbox{
\includegraphics[width=5cm]{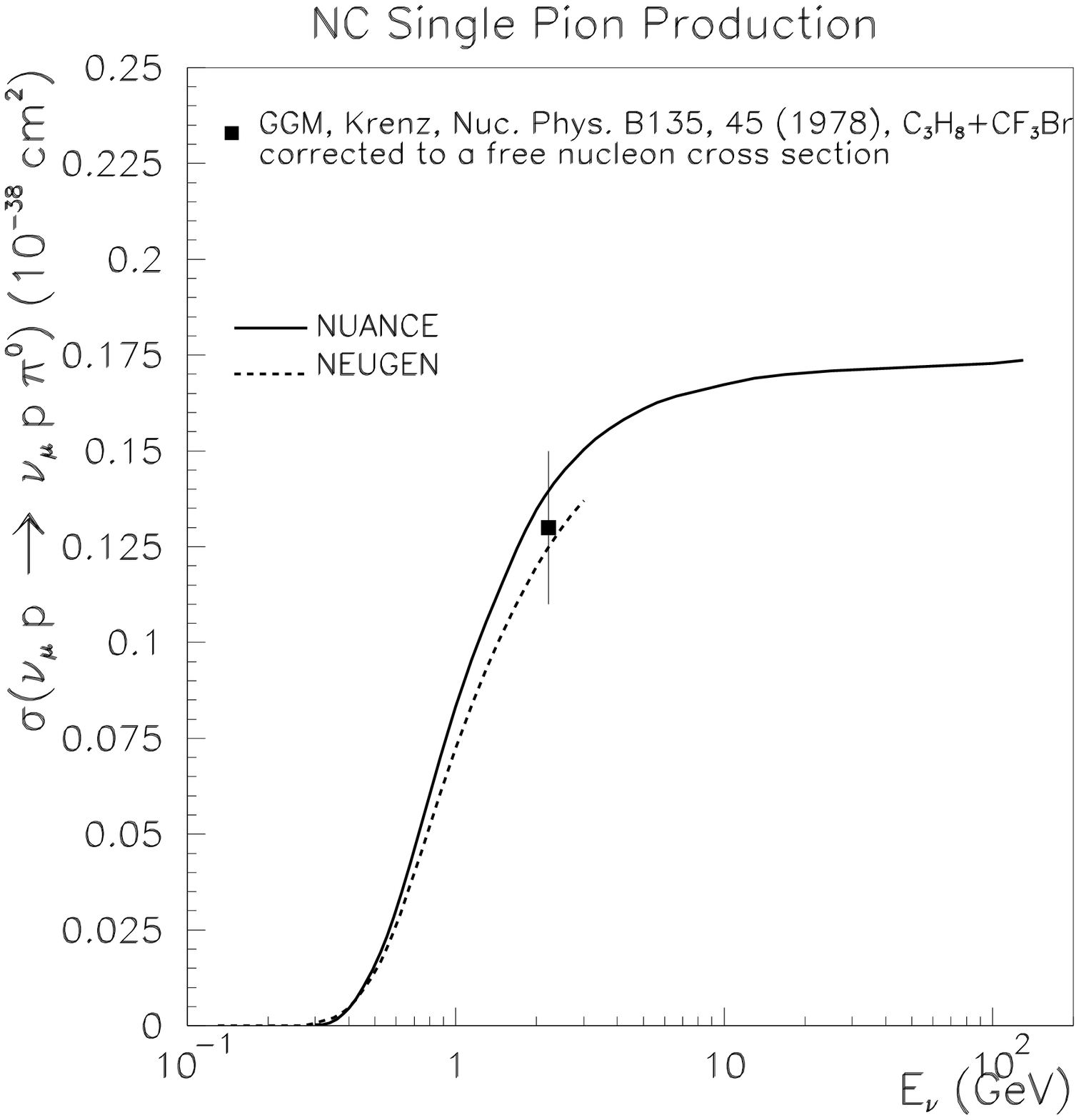}
\includegraphics[width=5cm]{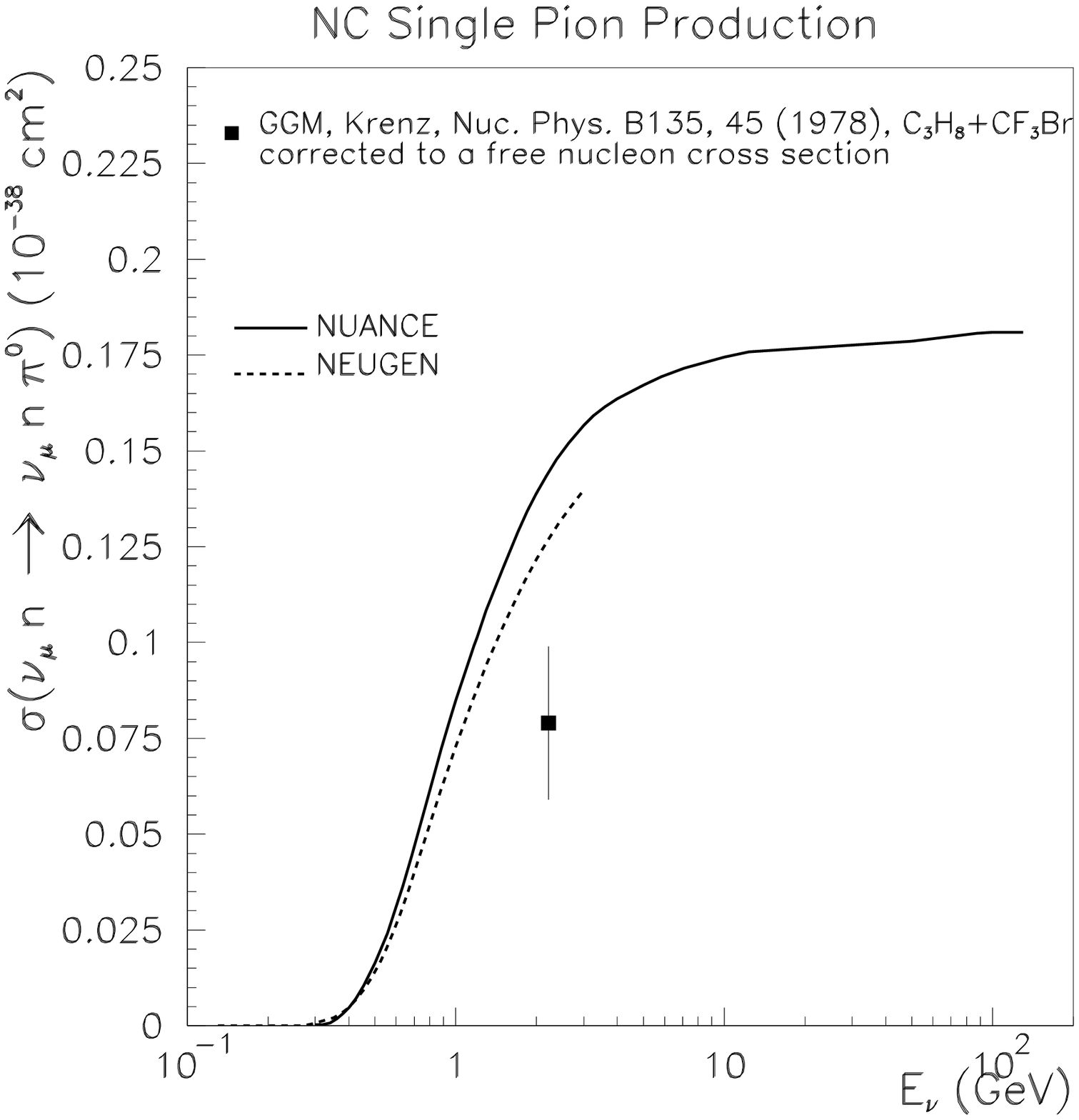}
}
\caption{Knowledge of single $\pi^0$
  neutrino production cross sections as a function of energy before K2K or
  MiniBooNE results (Zeller 2003).}
\label{fig:pi0-data}
\end{figure}

\begin{figure}
\centering
\includegraphics[width=9cm]{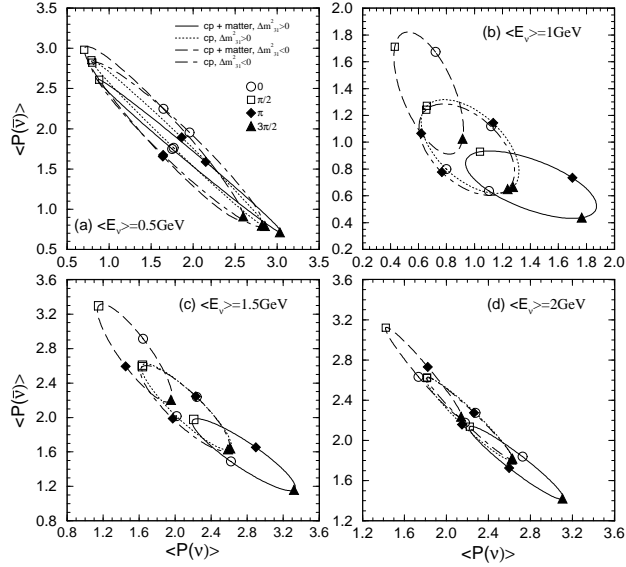}
\caption{The anti-neutrino vs. neutrino $\nu_\mu\to\nu_e$ transition
  probability in percent a baseline of $700$~km at different energies
  for different mass hierarchies and values of the CP violating phase
  $\delta$ (Minakata and Nunokata 2001).}
\label{fig:nue-physics}
\end{figure}

\index{electron neutrino appearance}
Because the $\nu_\mu\to\nu_e$ oscillation probability is so low,
the major impact on these appearance experiments, such as T2K and NOvA, is
expected to be from backgrounds to electron appearance.  The major such
background is the production of neutral pions which decay
into photons that shower and mimic electrons,
either because of a merging or loss of $\gamma$ rings in a Cerenkov
detector 
or because of the
merging or loss of one $\gamma$ in a calorimetric detector.
Unfortunately, this background is poorly constrained by existing data
(Figure~\ref{fig:pi0-data}).  The challenge becomes apparent when
looking at the precision needed for the physics goals of these
experiments.  Ultimately, as illustrated in
Figure~\ref{fig:nue-physics}, the transition probabilities will need to
be measured with sub-percent precision to measure the effect of CP
violation in neutrinos.  This places strict requirements on the understanding of
$\nu_e$ backgrounds in both neutrino and anti-neutrino beams.

This interest in neutrino interactions in the $0.5$ to $5$~GeV energy
region has led to the proposal and construction of a number of
dedicated neutrino cross section experiments designed to make these
measurements.  The K2K experiment recently built a near detector, ``SciBar'',
designed for such measurements which is currently running as
``SciBooNE'' in the Fermilab Booster neutrino beam.  The MINERvA
experiment is currently under construction for future operation in the
Fermilab NuMI beamline.

\section{Pointlike Interactions}

\index{neutrino interactions}
For a pedagogical explanation of neutrino cross section
phenomenology, it is helpful to start with the scattering of neutrinos
from effectively massless pointlike fermions, such as neutrino-electron scattering.
Although this interaction is of limited practical interest for
accelerator oscillation experiments, the calculation of pointlike
scattering will serve multiple purposes as we begin to explore more
complicated cross section phenomenology.  First, in the high energy
limit of neutrino-nucleon scattering, a good approximation to deep
inelastic scattering is to consider the scattering of neutrinos on
point-like quark constituents of the nucleus.  Second, the study of
scattering from pointlike particles will make a good point of
departure from which to study effects such as initial and final state masses and
the effect of structure in a target fermion.  Therefore, please
suspend skepticism of the usefulness of this particular exercise, 
and let us begin
to consider neutrino scattering on electrons.
\index{neutrino electron scattering}

The style of the lectures was to present examples illustrating the
phenomena of neutrino interactions.  Accordingly, what follows below
uses heuristic arguments and does not follow a style of rigorous
proof.  To paraphrase the humorist Michael Feldman, readers who are
sticklers for the whole truth should write their own lectures.
 
\subsection{Weak Interactions and Neutrinos}

The modern view of the weak interaction is not the four
fermion interaction of Fermi's theory, but rather an interaction
mediated by the exchange of massive $W$ and $Z$ bosons.  In the low
momentum limit, where the mediating boson is far off shell, the weak
interaction Hamiltonian governing the process $\nu f \to \ell/\nu+ f^\prime$ is 
\begin{equation}
{\cal H}_{weak}=\frac{4 G_F}{\sqrt{2}}
\left[ \bar{\ell}/\bar{\nu}\gamma_\mu\frac{(1-\gamma_5)}{2} \nu\right] 
\left[
\bar{f^\prime}\gamma^\mu\left(g_L\frac{1-\gamma_5}{2}+g_R\frac{1+\gamma_5}{2}\right)
f\right] +{\rm\textstyle h.c.} 
\end{equation}
where $f$, $f'$, $l$ and $\nu$ stand for an initial and final state fermion,
lepton and neutrino, respectively, 
$g_L$ and $g_R$ are the weak neutral-current chiral couplings, 
$\gamma_\mu$ are the standard Dirac matrices
and $\gamma_5\equiv i\gamma_0\gamma_1\gamma_2\gamma_3$.
Note that, like the Fermi theory, this form makes reference to a
single coupling constant, $G_F$, to which we shall return later.  It
does include an important component not recognized in the Fermi
theory, namely parity non-conservation. The factor $(1-\gamma_5)/2$ is
a projection operator onto left-handed states for fermions and
right-handed states for anti-fermions.

\begin{table}
\centering
\caption{Weak neutral-current couplings $g_L$ and $g_R$.}
\label{tab:nc-couplings}
\begin{tabular}{|c|c|c|}
\hline
$Z$ coupling & $g_L$  & $g_R$ \\ \hline
$\nu$ & $1/2$ & $0$ \\
$e$,$\mu$,$\tau$ & $-1/2+\sin^2\theta_W$ & $\sin^2\theta_W$ \\
$u$,$c$,$t$ & $1/2-(2/3)\sin^2\theta_W$ & $-(2/3)\sin^2\theta_W$ \\
$d$,$s$,$b$ & $-1/2+(1/3)\sin^2\theta_W$ & $(1/3)\sin^2\theta_W$ \\
\hline
\end{tabular}
\end{table}

\index{$W$ and $Z$ bosons}
\index{charged current interaction}
\index{neutral current interaction}

The Hamiltonian above also has provision for a neutral-current
interaction, mediated by the $Z$, in which the neutrino remains a
neutrino, and a charged-current interaction, mediated by the $W$ in
which the neutrino becomes a charged lepton.  A neutrino, weak or
flavor, eigenstate, $\nu_e$, $\nu_\mu$ or $\nu_\tau$, is associated
with the production of a charged lepton of the same generation in the
charged-current weak interaction.  
The weak interaction is maximally parity-violating
in the charged-current interaction, selecting only
left-handed fermions, and therefore the right handed charged-current
couplings are zero.  However, in the case of the neutral
weak interaction, these couplings are given in terms of the
electromagnetic and weak couplings by the electroweak unification
theory and their values for each species of fermion are given in
Table~\ref{tab:nc-couplings}.  Note the right-handed neutrino has no
weak couplings, neither in the neutral nor the charged current, which
makes it unique among the fermions.
   
The rigorous definition of this ``handedness'', or chirality, is 
equivalent to the definition of the left-handed (right-handed) projection
operator, $(1\mp\gamma_5)/2$.  If a particle is massless, this
chirality is equivalent to its helicity, i.e. the projection of 
its spin $\sigma$ along the
direction of the particle, $\sigma\cdot\hat{p}$.  The Hamiltonian
above indicates that neutrinos produced or participating in weak
interactions will be entirely left-handed.  Since neutrinos do have
mass, $m_\nu$, this implies that while the neutrino will primarily be negative
helicity, there will be a small positive helicity component,
frame-dependent, and proportional to $m_\nu/E_\nu$ 
where $E_\nu$ is the neutrino energy.  For most practical
purposes, this positive helicity component can be entirely neglected.

\begin{figure}
\centering
\includegraphics[width=7cm]{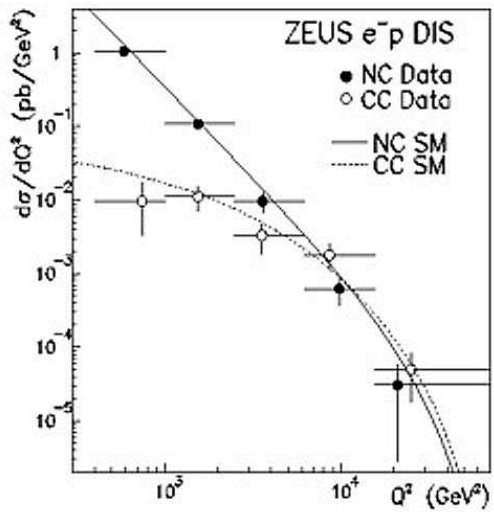}
\caption{Similarity of the strength of weak and electromagnetic
  interactions at high momentum transfer as illustrated by
  measurements of neutral and charged-current $ep$ scattering measured
  by the ZEUS experiment at HERA.}
\label{fig:ewk-q2}
\end{figure}

The final aspect of this form of the weak interaction to be explained
is the Fermi constant itself,
$G_F\approx1.166\times10^{-5}$~GeV$^{-2}$.  The dimensions and size of
the Fermi constant, which make the weak interaction ``weak'' at low
energies, have their origin in the propagator associated with the
exchange of the $W$ boson.  For a two body massless weak scattering
process,
\begin{equation}
\frac{d\sigma}{dq^2}\propto\frac{1}{(q^2-M_W^2)^2},
\label{eqn:prop}
\end{equation} 
where $M_W$ and $q$ are the mass of and the four-momentum carried by 
the $W$ boson\footnote{%
Suffice it to say that some details are glossed over
  in this statement.  It is rigorously true for a neutrino impinging
  on a target at rest in the lab that $d\sigma/dq^2=|{\cal M}|^2/(64\pi
  p_\nu^2 M_T^2)$ where $M_T$ is the mass of the target and ${\cal M}$
  is the matrix element of the scattering process.  There are many
  steps between this statement and the assertion above that the
  propagator ``factors'' out as shown above.}.  
For $|q^2|\ll M_W^2$, this propagator term gives a factor of
$M_W^{-4}$.  In the case of the electromagnetic interaction, this same
term becomes $q^{-4}$ since the mass of the exchanged boson, the
photon, is zero.  Figure~\ref{fig:ewk-q2} shows cross sections
of the neutral current process, $e^-p\to e^-p$, which has
contributions from both $\gamma$ and $Z^0$ exchange, and the charged
current process, $e^- p\to\nu X$, which is purely weak.  In these
processes, $q^2<0$, and we usually write $Q^2=-q^2$ by convention. We
see that when $Q^2<M_W^2$, the neutral current cross section is
rapidly falling with $Q^2$, while the charged-current cross section is
roughly constant.  However, beginning at $Q^2\sim M_W^2$, both
cross sections are roughly comparable and falling steeply with
$Q^2$. In the electroweak theory, $G_F$ can be expressed in terms
of $M_W$ and 
an overall weak coupling constant, $g_W$, as
\begin{equation}
G_F=\frac{\sqrt{2}}{8}\left( \frac{g_W}{M_W}\right) ^2;
\end{equation}
therefore, $g_W$ is a coupling of ${\cal O}(1)$ and roughly the same size
as the electromagnetic coupling constant in this unified theory of the
two interactions.

\subsection{Neutrino Electron Scattering}
\index{neutrino electron scattering}

\begin{figure}
\centering
\includegraphics[width=7cm]{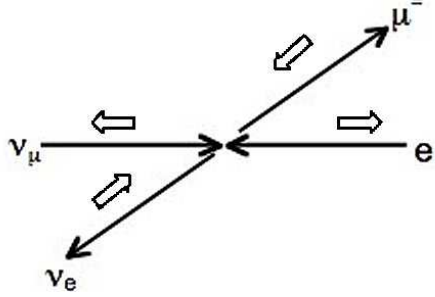}
\caption{Helicity in the massless limit of neutrino-electron
  scattering.}
\label{fig:nue-helicity}
\end{figure}

With this background, we are ready to calculate a neutrino-electron
scattering cross section.  For pedagogical simplicity, first consider $\nu_\mu
e^-\to\mu^-\nu_e$ at sufficiently high energies so that we may neglect all
masses in the problem, including the mass of the final state muon, but
not at such high energies that we need worry about the effect of $Q^2$
on the propagator $1/(Q^2+M_W^2)^2$.  In
this limit, chirality is equivalent to helicity.  
In the center-of-mass frame 
we can easily see that the two left-handed and negative
helicity particles in the final state have a total spin along the
interaction axis of $J_z=0$ (Figure~\ref{fig:nue-helicity}), and
therefore there is no preferred center-of-mass scattering angle.
Thus, 
\begin{equation}
\sigma=\int_0^{Q^2_{MAX}}
dQ^2\frac{d\sigma}{dQ^2}\propto\int_0^{Q^2_{MAX}}
dQ^2\frac{1}{(Q^2+M_W^2)^2}=\frac{Q^2_{MAX}}{M_W^4}.
\label{eqn:inv-muon-decay}
\end{equation}
That's it!  To complete the evaluation of the cross section, we need
only find the constant of proportionality, which turns out to be
$g_W^4/32\pi=M_W^4\times G_F^2/\pi$, and the maximum $Q^2$ that can be
exchanged.  Here $Q^2$, the negative of the square of the
four-momentum carried by the $W$ boson, is
$-(\underline{e}-\underline{\nu}_e)^2$, where the
\underline{underlined} terms represent four-vectors.  It is simple to
show that, in terms of $E^*_\nu$ and $\theta^*$  
the center-of-mass energy and scattering angle, 
$Q^2=2{E^*_\nu}^2(1-\cos\theta^*)$.
This means that $Q^2$
ranges between $0$ and $4{E^*_\nu}^2=s$
where $\sqrt{s}$ is the total available center-of-mass energy.
The cross section is
therefore
\begin{equation}
\sigma=\frac{G_F^2 s}{\pi}.
\end{equation} Numerically, this turns out to be
$\sigma = 17.2\times10^{-42}$~cm$^2 \times E_\nu$/GeV.
The proportionality to
the neutrino energy in the lab frame comes about computationally because, 
if the target electron is at rest, 
$s=m_e^2+2m_eE_\nu$
and the $m_e^2$ term can be neglected for neutrino beam energies of interest.
More
fundamentally, this proportionality to energy is a generic feature of
pointlike neutrino scattering at $Q^2$ below $M_W^2$ squared, since
$d\sigma/dQ^2$ is constant.

We now consider another process, this time the neutral current elastic
scattering $\nu_\mu e^-\to\nu_\mu e^-$.  How is this process different
from our previous example under the same energy approximations?
First, the process is a neutral current one, and therefore, as can be
seen in Table~\ref{tab:nc-couplings}, the interaction couples to both
the left-handed and right-handed electron.  In the massless left-handed case,
as before, the total spin along the interaction axis is $0$, but in
the right-handed case, the total spin along this axis is $1$.  The
right-handed case therefore differs from the case shown in
Figure~\ref{fig:nue-helicity} because if the target electron and
the outgoing lepton spins are flipped, there is a preference for forward
scattering as opposed to backward scattering due to conservation
of angular momentum along the interaction axis.  Therefore, while
$d\sigma/d\theta^*$ is constant for the left-handed target lepton, 
\begin{equation}
\frac{d\sigma_{J_z=1}}{d\theta^*} \propto \left(
\frac{1+\cos\theta^*}{2}\right) ^2
\label{eqn:helicity-suppressed}
\end{equation} 
for the right-handed target lepton.  Integrating this over all solid
angles leads to the conclusion that $\sigma_{J_z=1}=\sigma_{J_z=0}/3$,
where the reduced cross section can be understood from the suppression
of non-forward scattering due to the projection of spin from the
initial to the final state axes.

The couplings enter linearly into the matrix element and,
therefore, are squared in the cross sections.
With the effect of the initial state spin accounted for, we can write
\begin{eqnarray}
\sigma_{J_z=0}&=&\frac{G_F^2s}{\pi}\left(
-\frac{1}{2}+\sin^2\theta_W\right) ^2 \nonumber \\
\sigma_{J_z=1}&=&\frac{1}{3}\frac{G_F^2s}{\pi}\left(
\sin^2\theta_W\right) ^2 \nonumber \\
\sigma(\nu_\mu e^-\to\nu_\mu e^-)_{TOT}&=&\frac{G_F^2s}{\pi}\left(
  \frac{1}{4}-\sin^2\theta_W+\frac{4}{3}\sin^4\theta_W\right) .
\end{eqnarray}

Generalizing from the examples here, it's possible to derive all the
neutrino-electron elastic scattering processes in the massless limit.
Some, such as $\nu_e e^-\to \nu_e e^-$ have the added complexity of
both neutral and charged-current contributions.  In the case of this
reaction, the analysis is the same as that for $\nu_\mu e^-\to\nu_\mu
e^-$ scattering above with one exception.  The charged-current gives
an additional process contributing to the scattering from left-handed
electrons.  Because these processes have identical initial and final
states, they interfere, and therefore are correctly computed by adding
amplitudes rather than cross sections.  This leads to an effective
left-handed coupling for the process of $-1/2+g_L=-1+\sin^2\theta_W$,
where the $-1/2$ term represents the coupling of the charged-current
to the left-handed electron.  This results in a cross section of
\begin{equation}
\sigma(\nu_e e^-\to\nu_e e^-)_{TOT}=\frac{G_F^2s}{\pi}\left(
  1-2\sin^2\theta_W+\frac{4}{3}\sin^4\theta_W\right) ,
\end{equation}
much larger than that of the neutral-current only
process.
\index{cross section!neutrino electron scattering}
 
\subsection{The Effect of Initial and Final State Masses}

To this point, we have neglected the effect of massive particles.
This is not always a reasonable approximation, as it is simple to
illustrate with our initial example, $\nu_\mu
e^-\to\mu^-\nu_e$.  In the lab frame with a stationary target
electron, the total center-of-mass energy squared,
$s=m_e^2+2m_eE_\nu$.  However, in order to produce a muon in the final
state at all, $s\ge m_\mu^2$.  Solving, we find that this reaction only
occurs at all when the neutrino energy
\begin{equation}
E_\nu>\frac{m_\mu^2-m_e^2}{2m_e} ,
\end{equation}
which is approximately $11$~GeV.  Therefore, for practical cases such
as this one, we need a way to account for the effect of the final
state mass.

Recall that in our original derivation of this cross section in
Equation~\ref{eqn:inv-muon-decay}, we noted that we had to integrate
the roughly constant differential cross section $d\sigma/dQ^2$ over
the range of available $Q^2$ from zero up to a maximum value.  
In the massless limit, the range of $Q^2$ is, in fact, $0$ to $s$ as
asserted above.  In the presence of initial and final state masses,
these limits are more complicated:
\begin{eqnarray}
Q^2_{MAX,MIN}&=&(p^*_\nu\pm p^*_\mu)^2-\frac{(m_\mu^2-m_e^2)^2}{4s}
\nonumber\\
\Rightarrow Q^2_{MAX}-Q^2_{MIN}&=&4p^*_\nu p^*_\mu \nonumber
\nonumber\\
&\approx& s\left(1-\frac{m_\mu^2}{s}\right) \left(1+{\cal O}\left(
\frac{m_e^2}{m_\mu^2}\right) \right) .
\label{eqn:mass-sup}
\end{eqnarray}
In summary, the process is suppressed relative to its massless
cross section by a factor of $1-m_\mu^2/s$.  This suppression is
a factor that, while not general, recurs often in calculations of mass
suppression due to a single massless particle in the final state.
\index{cross section!mass effects}

Now consider a more complicated case of the inverse beta-decay
reaction in which reactor neutrinos were discovered, $\bar{\nu}_e p\to
e^+n$.  Here {\em both} particles in the final state are heavier than
their initial state counterparts: $m_e\approx 0.5$~MeV and
$m_n-m_p\approx 1.3$~MeV.  We can calculate the threshold energy, 
$E_\nu^{MIN}$, of the
reaction by observing that the heavy nucleon in the final state will
have zero kinetic energy to zeroth order in $m_e/m_n$.  Equating the
initial and final state $s$ under this condition, we find
\begin{equation}
E_\nu^{MIN}\approx \frac{(m_n+m_e)^2-m_p^2}{2m_p},
\end{equation}
which is approximately $1.8$~MeV.  If we define $\delta E\equiv
E_\nu-E_\nu^{MIN}$, we can then write
\begin{eqnarray}
s&=&(\underline{\nu}+\underline{p}) \nonumber\\
&=&m_p^2+2m_p(\delta E+E_\nu^{MIN}) \nonumber\\
&=&2m_p\times\delta E+(m_n+m_e)^2. 
\end{eqnarray}
Then the mass suppression factor is
\begin{eqnarray}
\xi_{mass}\equiv1-\frac{m_{final}^2}{s}&=&\frac{2m_p\times\delta
  E}{(m_n+m_e)^2+2m_p\times\delta E} \nonumber \\
&\approx & 
\left\{ 
\begin{array}{ll}
\delta E\times\frac{2m_p}{(m_n+m_e)^2}&{\rm\textstyle if}~\delta E\ll
m_p\\
1-\frac{(m_n+m_e)^2}{2m_p\times\delta E}&{\rm\textstyle
  if}~\delta E\gg m_p\\
\end{array}
\right. 
\label{eqn:imb-mass}
\end{eqnarray}
Note that for $\delta E\ll m_p$, the mass suppression $\xi_{mass}$ is linear in
$\delta E$, and therefore near threshold the cross section will
increase quadratically: one power from the $\delta E$ dependence in
Equation~\ref{eqn:imb-mass} and one power from the linear increase in
cross section with energy from pointlike scattering.

\section{Beyond Pointlike Scattering}

The astute reader will note, however, that I have yet to write down a
cross section for inverse beta decay because we are still missing a
key ingredient to do so.  Although electrons are pointlike, the
protons of inverse beta decay most certainly are not.  In the next
section of this lecture, we will consider the effect of the structure of
the target on neutrino interactions.

\subsection{Target Structure in \protect{$\nu N$} Elastic Scattering}

\index{elastic neutrino nucleon scattering}
To begin our exploration of scattering from pointlike scattering,
we will continue our investigation of inverse beta decay, $\bar{\nu}_e p\to
e^+n$.  This reaction is termed ``quasi-elastic''\footnote{Beware of
  nuclear physicists using the term ``quasi-elastic''.  It is also used to indicate
  nuclear dissociation in electromagnetic interactions, such as $e^- d
  \to e^- p n$.} 
in the sense that the target nucleon remains a single nucleon in the
final state and only changes its charge in the charged-current weak
interaction.  

\begin{figure}
\centering
\includegraphics[width=7cm]{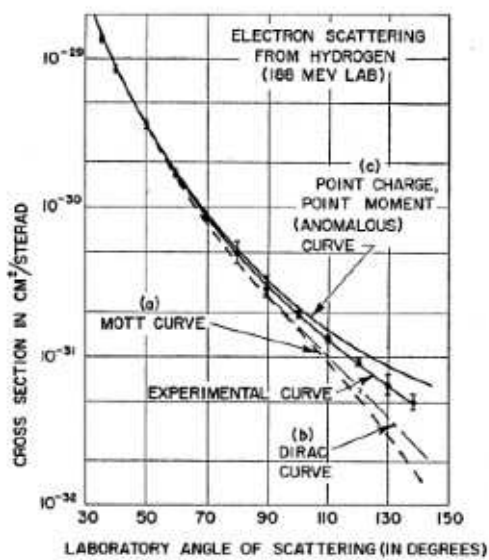}
\caption{Angular dependence of cross sections for scattering of
  $188$~MeV electrons.  The data measure the proton charge radius to
  be $(0.7\pm0.2)\times10^{-15}$~m. (McAlister and Hofstadter 1956).}
\label{fig:proton-structure}
\end{figure}

The target proton differs from an electron in several important
respects.  The couplings of composite particles like the proton are
not predicted by the electroweak theory, nor is the anomalous magnetic
moment, $(g-2)/2$, necessarily small for a composite particle.
Finally, the weak couplings may have a dependence on $Q^2$ which
reflects the finite size of the particle.
Figure~\ref{fig:proton-structure} shows data from some of the original
measurements of proton structure in $\sim 200$~MeV electron
scattering.  The increase in angle corresponds to an increase in the
$Q^2$ of the electromagnetic interaction.  As can be seen, the
experimental data do not agree with the prediction of the proton as a
Dirac particle, but require not only anomalous magnetic moment but also
finite-sized charge distribution to explain the suppression at
high $Q^2$ relative to a point charge.

\index{cross section!inverse beta decay}
The full cross section for inverse beta decay is
\begin{equation}
\sigma(\nub_e p\to e^+n)=\frac{G_Fs}{\pi}\times\cos^2\theta_C\times\xi_{mass}\times\left(
g_V^2+3g_A^2\right) ,
\label{eqn:ibd}
\end{equation}
where the first term is the point-like scattering cross section result
we derived for neutrino-electron scattering, the $\theta_C$
term takes into account the charged-current quark mixing transition
from a $u$ quark to a $d$ quark, $\xi_{mass}$ is defined in
Equation~\ref{eqn:imb-mass}, and $g_V$ and $g_A$ are the proton form
factors.  As mentioned above, the proton form factors and the relevant
(small) momentum transfer for this process at low energy are not
predicted by the electroweak theory, and must be experimentally
determined.  $g_V$ at low momentum transfer is the electric charge of
the proton, +1, and $g_A$ is determined by the neutron lifetime to be
$-1.26$.

\subsection{Deep Inelastic Scattering}
\label{sect:DIS}

Of course, another difference between a strongly bound target, such as
a nucleus, and an electron is that the strongly bound target can be
broken apart in the final state to create different particles.  In
such a case, what do we qualitatively expect to happen to the
cross section?

\begin{figure}
\centering
\includegraphics[width=9cm]{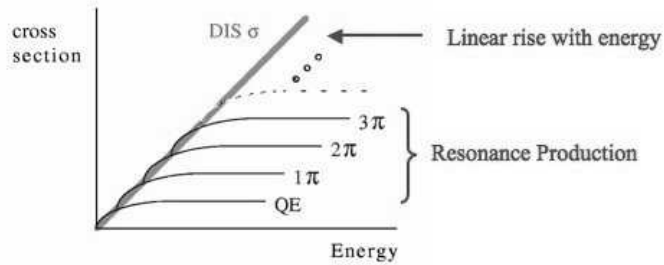}
\caption{A schematic diagram representing the rise of the
  cross section with energy as inelastic channels open up with energy.}
\label{fig:energy-rise-schematic}
\end{figure}

\index{deep inelastic neutrino scattering}
Consider first the elastic scattering process of neutrinos on
nucleons.  This total cross section will rise linearly with energy
when the energy is sufficiently low.  However, if the $Q^2$ of the
reaction is high enough, the differential elastic cross section,
$d\sigma/dQ^2$ will start to fall with $Q^2$ because the nucleon will
break apart when too much $Q^2$ is transferred.  At some point, the
cross section no longer rises with energy because the elastic process
only occurs up to a finite $Q^2$, and the $s$ at this high energy
exceeds that $Q^2$.  However, at the same point, new inelastic
processes, such as the production of a single pion will become
energetically possible.  These will rise with energy, initially
quadratically and then linearly until they too reach their $Q^2$
limit, at which point their cross section stops rising with energy.  As
illustrated in Figure~\ref{fig:energy-rise-schematic}, this process
repeats itself, resulting in a linear rise of the total cross section
with energy.

Of course, a linear rise with energy is exactly what is expected in the
case of point like scattering.  The picture above, while possibly
helpful in the region of transition between elastic and inelastic to
be discussed in Section~\ref{sect:gev}, is awkward for understanding
the high energy behavior of inelastic scattering.  Instead, we model
this process as the deep inelastic scattering of neutrinos from {\em
  quarks} inside the strongly bound system.  These quarks are
fundamental particles, and therefore the cross section of neutrino
quark scattering will rise linearly with energy.

\begin{figure}
\centering
\includegraphics[width=7cm]{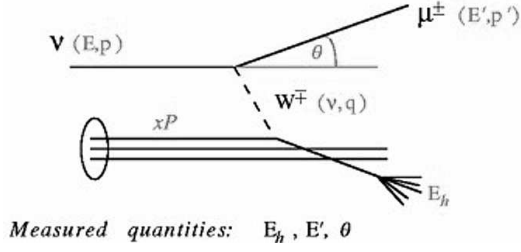}
\caption{Kinematic quantities in deep inelastic scattering.}
\label{fig:dis-kinem}
\end{figure}

We first need a common language of kinematics that is relevant for an
inelastic process such as $\nu N\to \ell X$ or its neutral current
counterpart, $\nu N\to\nu X$.  As shown in Figure~\ref{fig:dis-kinem},
we define the energy and four-momentum in the lab frame of the
incoming neutrino, the outgoing lepton and the weak boson,
respectively, to be $p = (E,p)$, $p^\prime = (E^\prime,p^\prime)$ 
and $q = (\nu,q)$, and
we also define the lab scattering angle of the outgoing lepton as
$\theta$, the four-momentum of the target as $P$, and the energy of
the hadronic recoil in the lab frame as $E_h$.  As before we define
the negative of the $W$ four-momentum squared
\begin{equation}
Q^2\equiv-q^2=-(p^\prime-p)^2\approx 4EE^\prime\sin^2(\theta/2). 
\end{equation}
Note that this definition is given purely in terms of variables on the
well-defined leptonic side of the event.  We will follow this
convention as much as possible, expressing quantities in the lab in
terms of leptonic variables and the initial target mass, $M_T$.  We
may define other invariants, such as the lab energy transfer, $\nu$,
the inelasticity, $y$, and the Feynman scaling variable, $x$:
\begin{eqnarray}
\nu&\equiv&\frac{q\cdot P}{\sqrt{P\cdot P}}=E-E^\prime, \nonumber \\
y&\equiv&\frac{q\cdot P}{p\cdot P}=\frac{\nu}{E},\nonumber \\
x&\equiv&\frac{-q\cdot q}{2(p\cdot q)}=\frac{Q^2}{2M_T\nu}.
\label{eqn:dis-kinem}
\end{eqnarray}
The center-of-mass scattering energy, $\sqrt{s}$, and the mass of the
hadronic recoil system, $W$, can also be written in term of leptonic
variables $x$, $y$ and $\nu$ and the target mass $M_T$:
\begin{eqnarray}
s&\equiv&(p+P)^2=M_T^2+\frac{Q^2}{xy}, \nonumber \\
W^2&\equiv&(q+P)^2=M_T^2+2M_T\nu-Q^2.
\end{eqnarray}

In the picture of neutrinos scattering from constituents of strongly
bound systems, the ``parton'' interpretation of deep inelastic
scattering, the variable $x$ has a special interpretation as the
fractional momentum of the target nucleon carried by the parton in a
frame where the target momentum is very large.  The common picture of
this frame is that the nucleon, as seen by the incoming lepton, is
flat and static because of length contraction and time dilation, and
the incoming lepton interactions with a single one of these frozen
partons, carrying a momentum fraction $x$.  In this picture, we can
define effective masses for the initial and final state partons,
\begin{eqnarray}
m_q^2&=&x^2P^2=x^2M_T^2, \nonumber\\
m_{q^\prime}^2&=&(xP+q)^2
\end{eqnarray}

\index{parton distribution functions}
\index{partons}
To make sense of deep inelastic scattering, we cannot merely consider
the hard process of neutrinos scattering from quarks; we must also
place those quarks inside the target hadron.  This is made possible by
the Factorization Theorems of QCD which allow us to write a scattering
cross section
for a hadronic process in terms of cross sections for scattering from
partons convoluted with a parton distribution function $q_h(x)$:
\begin{equation}
\sigma(\nu+h\to\ell+X)=\sum_q \int dx~ \sigma(\nu+q(x)\to\ell+X)q_h(x).
\end{equation}
The parton distributions, while not (yet) something we can calculate
from principles of QCD, are universal.  Therefore, they can be
determined in one process and applied to another process.

\begin{figure}
\centering
\includegraphics[width=8cm]{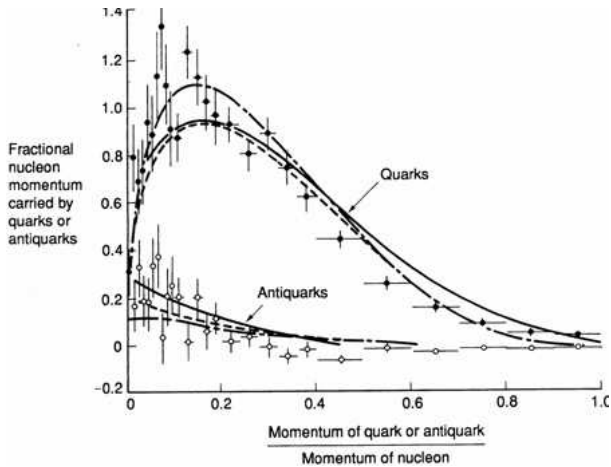}
\caption{Distribution of quark and anti-quark momentum density in the
nucleon as a function of $x$.}
\label{fig:pdf-x}
\end{figure}

Figure~\ref{fig:pdf-x} shows an illustration of typical quark and
anti-quark distributions in a nucleon at moderate $Q^2$.  The parton
distribution function (PDF), $q(x)$ gives the number density of quarks
of a given $x$.  If quarks, carried all the momentum of the nucleon,
$\int xq(x)dx=1$; however, in reality this integral is significantly
less than one.  This momentum sum also turns out to be logarithmically
dependent on $Q^2$, as are the PDFs more generally.  These slow
changes with $Q^2$ are called ``scaling violation'', in reference to
the Feynman scaling variable, $x$.  They result from the strong
interactions of the quarks themselves in the nucleon.  There is a
duality between $Q^2$ and distance scales, with higher $Q^2$
interactions probing features at small distance scales.  At these
small scales, the strong interactions among partons in the nucleon
will cause quarks to radiate gluons and gluons to split into quarks
and anti-quarks.  The net results, whose effects have been calculated
quantitatively in perturbative QCD, are that quarks and anti-quarks
increase in number at higher $Q^2$, but their average fractional
momentum decreases\footnote{This and other topics in perturbative QCD
make for fascinating exploration in detail, but are well beyond the
scope of these lectures.  I highly recommend the CTEQ Collaboration
Handbook of Perturbative QCD (Sterman et al 1995) for a pedagogical
introduction to these topics.}.

\subsubsection{Deep Inelastic Scattering as Elastic Neutrino-Quark Scattering}

Now that we have established the link between neutrino deep inelastic
scattering and elastic neutrino-quark scattering, we can apply what we
have learned about elastic scattering to deep inelastic scattering.
Recall that for the charged current $\nu q\to\ell q^\prime$ process,
we expect a cross section of $G_Fs/\pi$, up to a possible angular suppression
accounting for total spin in the initial and final state.  
But $s=M_T^2+2M_TE_\nu$ in the lab frame.  We have just
learned that for each quark, the initial state target mass is $xm_N$,
where $m_N$ is the nucleon mass, so the total effective target mass is
of the same order of magnitude as the nucleon mass.  Compared with the case of
neutrino-electron scattering where the target mass was $m_e$, the
cross section for deep inelastic scattering will be approximately three orders of
magnitude larger!  

We can also look at chirality and total spin in the reactions of
neutrinos and quarks.  Again, in the high energy limit where helicity
and chirality are equivalent, consider the center-of-mass frame as we
did in Figure~\ref{fig:nue-helicity}.  For the case of a quark, the
charged-current weak interaction will pick out left-handed quarks just
as it did left-handed electrons, and there will be no net spin along
the interaction axis.  By contrast, for the case of neutrino
scattering from anti-quarks, the target will be right-handed in the
center-of-mass frame, and there will be a net spin of $1$ along the
interaction axis.  As we argued in Equation~\ref{eqn:helicity-suppressed},
for the case of a right-handed target, the back-scattering is
suppressed, and the overall cross section is reduced by a factor of
three.  A convenient kinematic relationship exists between the
center-of-mass scattering angle, $\theta^*$ and the inelasticity, $y$,
\begin{equation}
\left( \frac{1+\cos\theta^*}{2}\right) = 1-y,
\end{equation}
and therefore, $d\sigma_{J_z=1}/d\theta^*\propto (1-y)^2$.

\begin{figure}
\centering
\includegraphics[width=7cm]{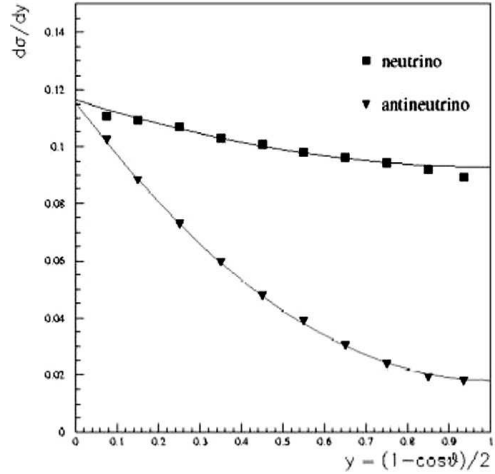}
\caption{Neutrino and anti-neutrino deep inelastic scattering
  cross sections as a function of $x$.}
\label{fig:dis-x}
\end{figure}

\index{cross section!deep inelastic neutrino scattering}
The same argument that leads to a back-scattering or high $y$
suppression of the cross section for neutrino-antiquark scattering
also holds for antineutrino-quark scattering, and similarly
antineutrino-antiquark scattering has no suppression.  Therefore
\begin{eqnarray}
\frac{d\sigma(\nu q)}{dxdy}&=
\frac{d\sigma(\nub \bar{q})}{dxdy}&\propto 1, \nonumber\\
\frac{d\sigma(\nub q)}{dxdy}&=
\frac{d\sigma(\nu \bar{q})}{dxdy}&\propto (1-y)^2.
\label{eqn:nu-q-bar-y}
\end{eqnarray}
This fact, combined with the smaller momentum fraction carried by
antiquarks than is carried by quarks (Figure~\ref{fig:pdf-x}), means that the total
anti-neutrino cross section is approximation factor of two smaller than
the neutrino cross section on nucleons.  Differential cross sections of
each are shown in Figure~\ref{fig:dis-x}.

\subsubsection{Structure Functions in Deep Inelastic Scattering}

\index{structure functions in deep inelastic scattering}
We have approached deep inelastic scattering both from its 
interpretation as neutrino-quark elastic scattering, and also 
by purely considering the kinematics.  Beyond kinematic constraints,
conservation laws and
Lorentz invariance also provide model independent constraints on
the possible forms of inelastic scattering cross section, and in this
picture, information about the structure of the target is contained in
a number of general ``structure functions''.  If we consider
the case of zero lepton mass, there are three structure functions that
can be used to describe the scattering, $2xF_1$, $F_2$ and $xF_3$:
\begin{equation}
\frac{d\sigma^{\nu,\nub}}{dxdy}\propto\left[
y^22xF_1(x,Q^2)+\left( 2-2y-\frac{M_Txy}{E}\right)
F_2(x,Q^2)\pm y(2-y)xF_3(x,Q^2)\right] .
\end{equation}
Note that $xF_3$ is a structure function that is not present in electromagnetic
interactions, and is only allowed because of the parity violation of
the weak interaction.

There is an approximate simplification with a
model of massless, free spin-$1/2$ partons, first derived by Callan
and Gross, $2xF_1=F_2$.  The Callan-Gross relation implies that the
intermediate boson is completely transverse, and so violations of
Callan-Gross are often parameterized by $R_L$, defined so that
\begin{equation}
R_L\equiv\frac{\sigma_L}{\sigma_T}=\frac{F_2}{2xF_1}\left(
1+\frac{4M_Tx^2}{Q^2}\right) .
\end{equation}
Contributions to $R_L$ arise because of processes internal to the
target, like gluon splitting $g\to q\bar{q}$ which are
calculable in perturbative QCD, and because of the target mass, $M_T$ 

Continuing with the assumptions of the validity of the Callan-Gross
relation and of massless targets, we can match the $y$ dependence of the
structure functions with the $y$ dependence of elastic scattering from
quarks and anti-quarks to make assignments of structure functions with
parton distributions.  In this limit, the coefficient in front of
$xF_3$ simplifies to $1-(1-y)^2$, and the coefficient multiplying $2xF_1=F_2$
is $1+(1-y)^2$.  From Equation~\ref{eqn:nu-q-bar-y}, the former would be
associated with the non-singlet contribution of $q-\qbar$ and the
later with the sum $q+\qbar$.  Furthermore, for the charged-current,
there is a charge selection, namely, a neutrino cannot produce a quark
or anti-quark
by sending its $W^+$ to a target quark unless that target quark has
negative charge; otherwise, the resulting final state would have to
have charge greater than $+1$ and would not be a quark.  Putting all
these constraints together, we find:
\begin{eqnarray}
2xF_1^{\nu p,~{\rmt CC}}&=&x\left[
  d_p(x)+\bar{u}_p(x)+s_p(x)+\bar{c}_p(x)\right], \nonumber\\
xF_3^{\nu p,~{\rmt CC}}&=&x\left[
  d_p(x)-\bar{u}_p(x)+s_p(x)-\bar{c}_p(x)\right], 
\end{eqnarray}
where $q_p(x)$ refers to the PDF of a given quark flavor in the proton
and where the contribution from third generation quarks, which have
very small PDFs, is neglected.

Just as with the neutrino-electron scattering, the neutral current
case is more complicated because the neutral current couples to
quarks of both helicities with a non-trivial coupling constant.  
However, unlike the charged-current case,
there is no selection based on quark charge.  The neutral current
structure functions under the same assumptions are:
\begin{eqnarray}
2xF_1^{\nu p,~{\rmt NC}}&=&x\left[
\left( u_L^2+u_R^2\right) \left(
u_p(x)+\bar{u}_p(x)+c_p(x)+\bar{c}_p(x)\right) \right.
\nonumber\\ 
&&\left. +\left( d_L^2+d_R^2\right) \left(
d_p(x)+\bar{d}_p(x)+s_p(x)+\bar{s}_p(x)\right) \right] ,
 \nonumber\\
xF_3^{\nu p,~{\rmt NC}}&=&x\left[
\left( u_L^2+u_R^2\right) \left(
u_p(x)-\bar{u}_p(x)+c_p(x)-\bar{c}_p(x)\right) \right. \nonumber\\ 
&&\left. +\left( d_L^2+d_R^2\right) \left(
d_p(x)-\bar{d}_p(x)+s_p(x)-\bar{s}_p(x)\right) \right] ,
\end{eqnarray}
where the new notation here, e.g., $u_{L,R}$, refers to the left and
right-handed neutral current couplings of up or down type quarks.

\index{parton distribution functions}
Some simplification in the case of the charged-current can be obtained
for the practical case where the target material consists of an
isoscalar nucleus with equal numbers of neutrons and protons.  The
light PDFs of the neutron are standardly assumed to be related to the
PDFs of the proton by isospin symmetry,
\begin{eqnarray}
u_p(x)&=&d_n(x), 
 \nonumber\\
d_p(x)&=&u_n(x), 
\end{eqnarray}
and the PDFs of the heavy quarks, $s(x)$ and $c(x)$ are assumed to be
identical in neutrons and protons and identical with their anti-quark
distributions since they result from gluon splitting and not the valence
quark content of the nucleon.  Under these assumptions,
\begin{eqnarray}
2xF_1^{\nu N,~{\rmt CC}}&=&x\left[
  u(x)+d(x)+\bar{u}(x)+\bar{d}(x)+2s(x)+2c(x)\right]\nonumber \\
&=&x(q(x)+\qbar(x)), \nonumber\\
xF_3^{\nu N,~{\rmt CC}}&=&x\left[
  u(x)+d(x)-\bar{u}(x)-\bar{d}(x)+2s(x)-2c(x)\right]\nonumber\\
&=&x(q_{val}(x)+2s(x)-2c(x))
\end{eqnarray}
where the PDFs written are those of the proton and where
$q_{val}(x)\equiv q(x)-\qbar(x)$.  Note the particularly simple forms,
these structure functions have, at least in the limit of neglecting
the heavy quarks.

\subsubsection{$\nu_\tau$ Charged Current Interactions}

\index{tau neutrino interactions}
A challenging endeavor to apply our theory of deep inelastic
scattering is $\nu_\tau$ appearance experiments such as
OPERA.  The full calculation of lepton mass effects is beyond the
scope of these lectures. Note that all that has preceded this,
including the definitions of the structure functions, assumed massless
leptons.  But again, we can apply our mass suppression formalism to
get an approximation of the effect.

\begin{figure}
\centering
\includegraphics[width=7cm]{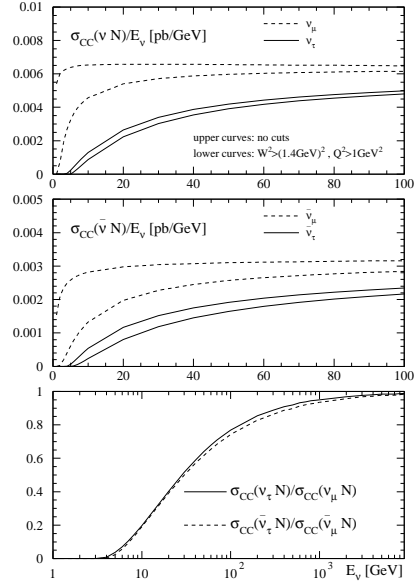}
\caption{Mass suppression of $\nu_\tau$ deep inelastic scattering
  cross sections (Kretzer and Reno 2002).}
\label{fig:tau-mass-dis}
\end{figure}

As we argued in Equation~\ref{eqn:mass-sup}, the generic form of the mass
suppression is $(1-m_{final}^2/s)$.  Since deep inelastic
scattering is neutrino-quark elastic scattering, 
the relevant quantity for $s$ here
is the $s$ of the neutrino-quark system, which is $\hat{s}=xs$.  The
form of the mass suppression for $\tau$ production from a given parton
$x$ is then $1-m_\tau^2/(xs)$.
This implies that at low $x$ the mass suppression will be large at
much higher energies than at high $x$, and thus qualitatively, the
rise of the cross section relative to muon neutrino charged current
scattering will be very slow.  This can be seen in the full calculation
illustrated in Figure~\ref{fig:tau-mass-dis}.

\subsubsection{Charm Production by Neutrinos}

\index{charm production}
Another way that massive final state corrections can enter into deep
inelastic scattering is the production of charm quarks in neutrino
deep inelastic scattering.  Although there are few charm quarks to be
found in the proton itself (after all, $m_c>m_p$), the sea has a large
number of strange quarks, roughly half as many as either of the light
quark species.  Since the Cabibbo-favored charged-current process
turns these strange quarks into charm quarks, production of charm
quarks is a significant fraction of the charged-current cross section.

Let's return to the kinematic variables of Equation~\ref{eqn:dis-kinem} to
study the effect of the final state quark mass.  Production of a charm
quark in the final state implies that $m_c^2=(q+\xi P)$, where $\xi$
represents the fractional momentum of the initiating quark instead of
the usual Feynman scaling variable $x$.  If $\xi\ll 1$, then
\begin{equation}
\xi\approx\frac{-q^2+m_c^2}{2P\cdot
  q}=\frac{Q^2+m_c^2}{2M_T\nu}=x\left( 1+\frac{m_c^2}{Q^2}\right) .
\end{equation}
The reason introducing $\xi$ as distinct from $x$ now becomes obvious.  
The $x$
variable as defined in terms of leptonic side variables is no longer
the same as the fractional momentum carried by the target quark, but
is in fact smaller.  Therefore, for a given set of scattering
kinematics, the initiating quark must carry a higher fractional
momentum and thus will be less common than in the case where a light
quark is produced in the final state.  This formalism for treating the
production of massive quarks is referred to as ``slow rescaling''.

\begin{figure}
\centering
\mbox{
\includegraphics[width=7cm]{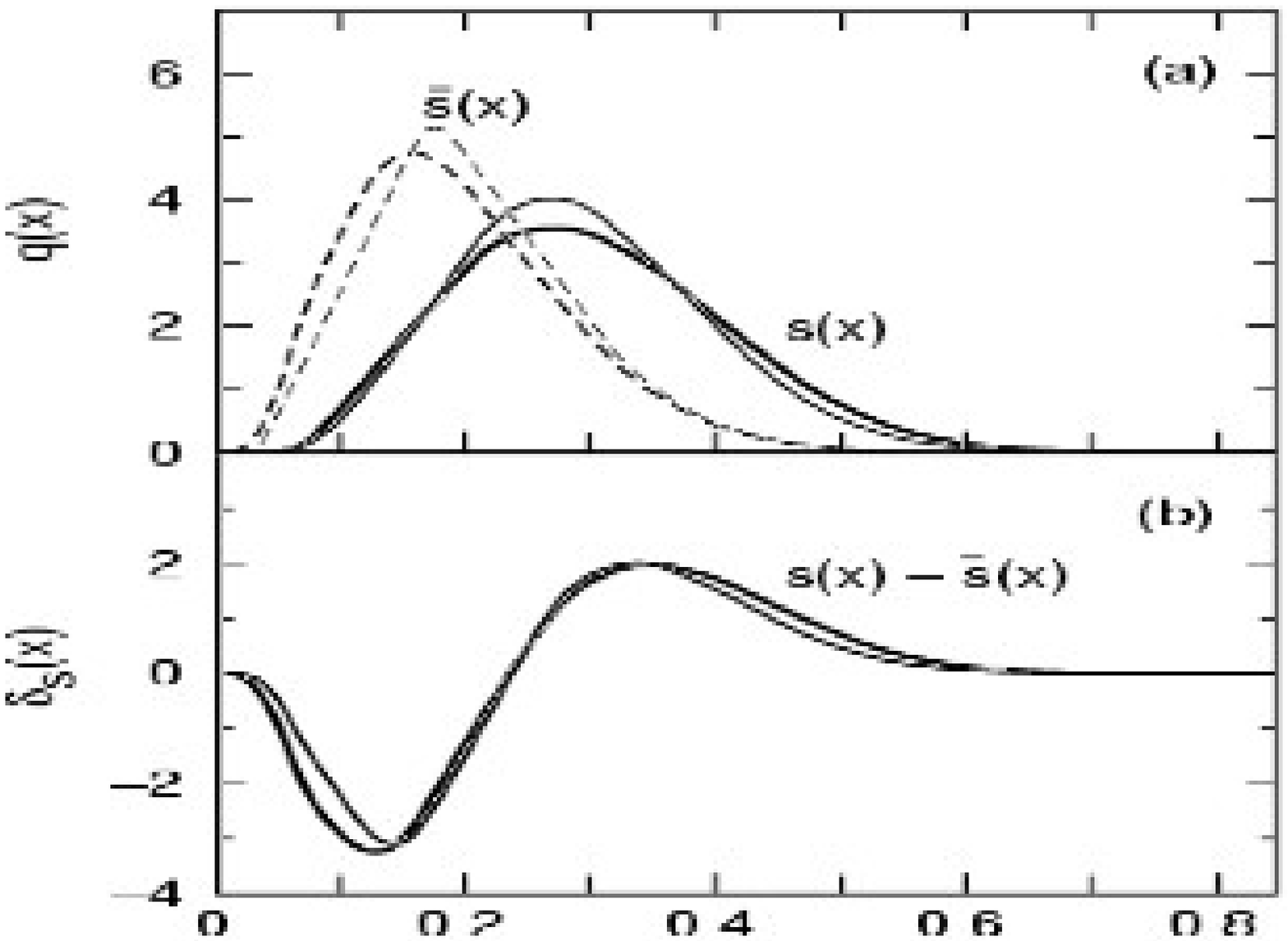}
\includegraphics[width=6cm]{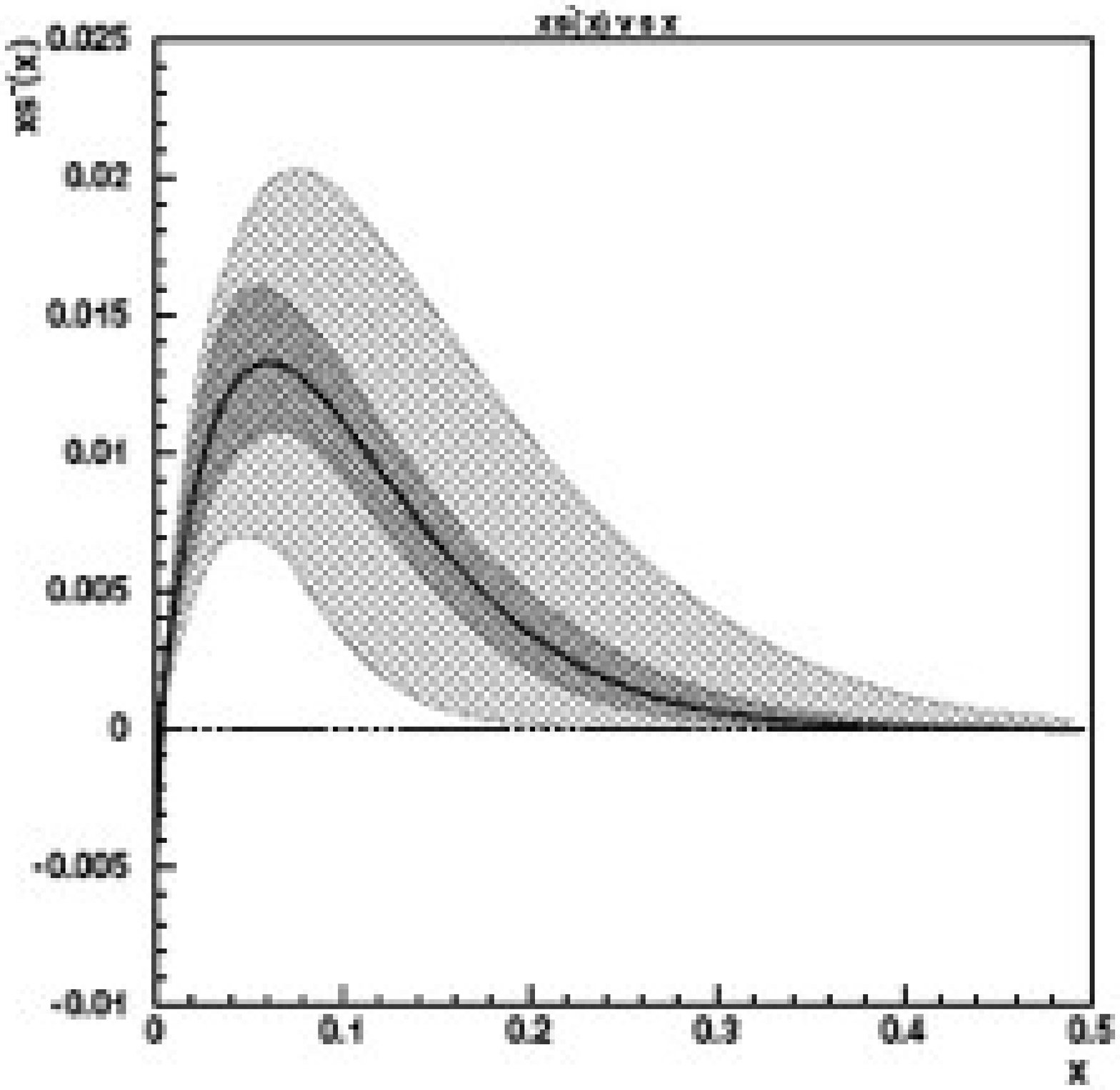}
}
\caption{Non-perturbative model of Brodsky and Ma 
  for $s(x)-\bar{s}(x)$ (left) compared
  with data from NuTeV (right).}
\label{fig:s-sbar}
\end{figure}

One of the best ways to actually measure the strange quark content of
the nucleon is to measure charged-current charm quark production tagged by the
semi-muonic decay of charm in a neutrino experiment.  In other words,
determine $s(x)$ by measuring 
$\nu_\mu+s\to\mu^-+c+X$, $c\to\mu^++X$ and its anti-neutrino analog,
each of which give two high momentum muons in the final state and are
commonly referred to as ``dimuon'' events.  

There is currently some debate about parton distributions regarding whether
the strange and anti-strange seas carry equal momentum.  Strange
quarks and anti-quarks generated by perturbative processes should be
nearly symmetric in momentum, but there are non-perturbative effects that can
lead to differences in their momentum.  The NuTeV experiment has
recently completed an analysis of dimuon events induced by neutrino
beams and anti-neutrino beams which therefore separately measure the
strange and anti-strange quark distributions in the nucleus.
Figure~\ref{fig:s-sbar} shows a comparison of one theoretical
prediction with the measurement of this momentum asymmetry from NuTeV.

\section{Transitions between Elastic and Inelastic Scattering}

\begin{figure}
\centering
\includegraphics[width=8cm]{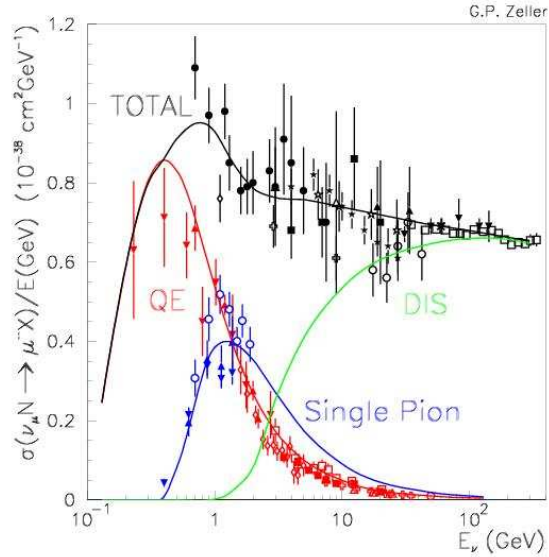} 

\caption{A compilation of neutrino cross sections, shown as
  $\sigma/E_\nu$, in the GeV region with quasi-elastic, deep inelastic
  and single pion cross sections shown separately.  (Figure courtesy G.P.~Zeller).}
\label{fig:gev-sigma}
\end{figure}

To this point, we have explored elastic scattering of neutrinos from pointlike
particles and a high energy limit of neutrinos scattering
inelastically from nucleons where the neutrino effectively scatters
from free quarks in the target.  If we look at the cross sections
shown in Figure~\ref{fig:gev-sigma}, we see both the elastic and deeply
inelastic cross sections co-existing over a broad region with a
significant component over nearly an order of magnitude in energy being the
``barely inelastic'' process of single pion production.  This
transition occurs at these energy values because the ``binding energy'' of
the the target nucleon is approximately $\lambda_{QCD}$, which is 
the scale of a typical momentum exchange for scattering of a
neutrino with $1$~GeV energy.

This section of the lectures will explore a few features of regions of
transition.  Because it is of the most interest for oscillations, we
will largely focus on the transition between nucleon elastic and
inelastic at neutrino energies near a GeV, but we will conclude with
comments on other transition regions of interest.

\subsection{The GeV Region}
\label{sect:gev}

We have exhaustively described the deep inelastic scattering limit of
Figure~\ref{fig:gev-sigma}, but have not spent much time describing
the quasi-elastic cross section, e.g. $\nu_\mu n\to \mu^- p$.  At low
energies, we expect by the same arguments given in other cases of
lepton mass, a suppression due to the muon mass going roughly as
\begin{eqnarray}
\frac{\sigma}{\sigma_{\rmt massless}}&\sim& 1-\frac{(M_T+m_\mu)^2}{s}
\nonumber \\ 
&=&1-\frac{M_T^2+2M_Tm_\mu+m_\mu^2}{M_T^2+2E_\nu M_T}
\nonumber\\ 
&\approx &\frac{2E_\nu}{M_T}~~~{\rmt if}~E_\nu<<M_T.
\end{eqnarray}  
Thus the cross section at low energies will be quadratic until
$E_\nu\sim m_T/2$, as shown in Figure~\ref{fig:gev-sigma}.  However, we also
see that the cross section stops growing above $E_\nu\sim 1$~GeV.  As
we observed in Section~\ref{sect:DIS}, above a
sufficiently high $Q^2$, $d\sigma/dQ^2$ begins to fall because
interactions at higher $Q^2$ tend to break apart the target nucleon
and therefore are not quasi-elastic.

\index{quasi-elastic neutrino nucleon scattering}
Nucleon structure also plays a significant role in quasi-elastic
scattering.  As with deep inelastic scattering, it is relatively
straightforward to write a cross section formula for quasi-elastic
scattering; however, in the end there are unknown form factors that
enter the calculation which must be determined experimentally.  The
cross section is usually parameterized in terms of vector, $F_V$, and axial
vector form factors, $F_A$,
\begin{equation}
F_{V,A}\approx\frac{F_{V,A}(0)}{\left( 1+Q^2/M_{V,A}^2\right) ^2},
\end{equation}
in the so-called ``dipole approximation'' (Llewellyn Smith 1972).
These are only phenomenological approximations to the true form
factors, and precise measurements of the  vector form factor $F_V$ in
electron scattering show significant deviations from the dipole form
at $Q^2\gg M_V^2$ where $M_V\approx0.84$~GeV.  The axial vector form
factor parameters are well determined only at $Q^2\approx 0$ (recall
the discussion following Equation~\ref{eqn:ibd}), and the best current estimates
from data of the axial mass give $M_A\approx 1.1$~GeV with significant
theoretical and experimental uncertainties.
\index{form factors}

\begin{figure}
\centering
\includegraphics[width=8cm]{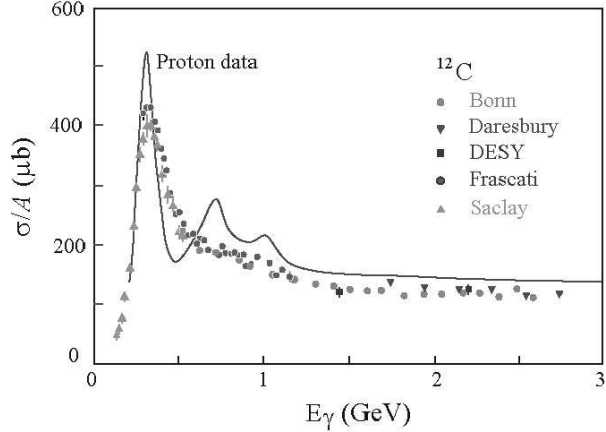}
\caption{Photo-absorption data on protons (line) and nuclei (data
  points) as a function of energy illustrating the effect of Fermi
  smearing on resonance production.}
\label{fig:nuclear-smear}
\end{figure}

As the cross section becomes barely inelastic, this region is often
called the ``resonance region'' because it is dominated by the
production of discrete baryon resonances in the final state. Recall
that the mass of the hadronic system, $W$, is given by
$W^2=M_T^2+2M_T\nu(1-x)$.  In the barely inelastic regime, this cannot
take any arbitrary value because there must be a baryonic state
available at that mass.  As the solid line in
Figure~\ref{fig:nuclear-smear} illustrates in a different process,
photo-nuclear absorption, there are discrete excitation lines
corresponding to specific broad baryon resonances.  The lowest mass
excited baryonic state is the $\Delta(1232)$ resonance
which is very visible and separated as the first peak in
Figure~\ref{fig:nuclear-smear}.  Above
the $\Delta(1232)$, resonances tend to overlap one another and
approach a continuum.

How is it possible to understand such a complicated set of overlapping
final states?  One way to gain a good qualitative and the beginning of
a quantitative understanding is through Bloom-Gilman duality.  The
ideal of duality is that, on average, one can model the behavior of
discrete hadronic states through the behavior of their underlying
quark content.  This emperically successful idea straddles the border
between asymptotically free and confined states in QCD.  

\begin{figure}
\centering
\includegraphics[width=9cm]{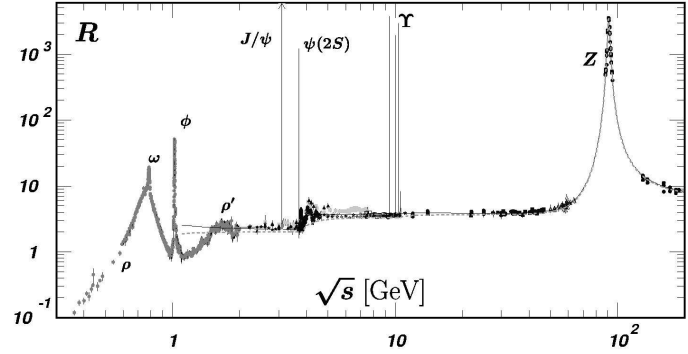}
\caption{The ratio of $e^+e^-$ annihilation cross section into hadrons
  divided by that into muons}.
\label{fig:bloom-gilman}
\end{figure}

\begin{figure}
\centering
\includegraphics[width=8cm]{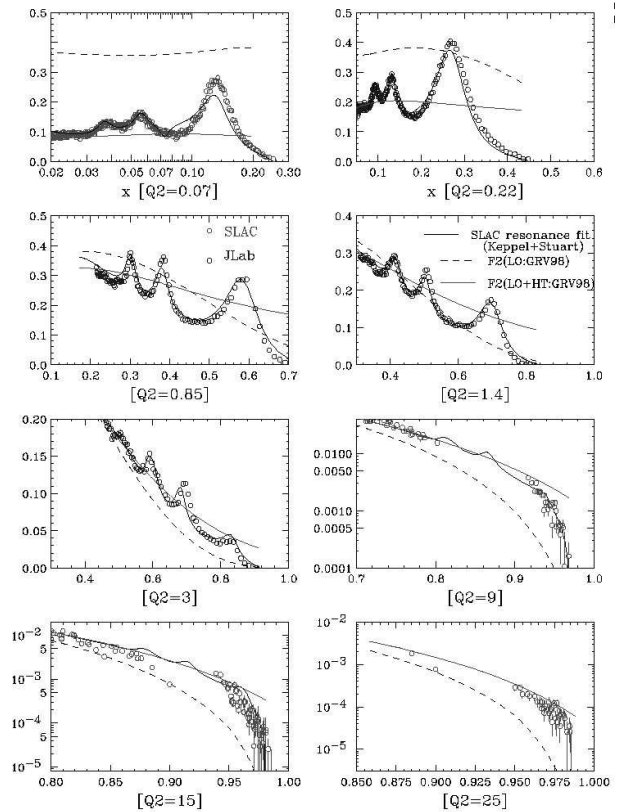}
\caption{Electron scattering data in the baryon resonance region
  compared to a quark-model calculation (Bodek and Yang 2002).}
\label{fig:bodek-yang}
\end{figure}

The most famous example of Bloom-Gilman duality is illustrated in the
quantity $R=\sigma(e^+e^-\to$~hadrons$)/\sigma(e^+e^-\to\mu^+\mu^-)$.
Shown in Figure~\ref{fig:bloom-gilman} is $R$ compared against a
prediction from the quark model that $R=N_C\sum Q_q^2$, where 
$Q_q$ and $N_C$ are the charge and number of color states for the quark, 
respectively.
The sum
runs over all quark states that can be produced at a given $s$.  This
method works well to describe the cross-section over a complicated mix
of final states that can be found 
well above the $s\bar{s}$ ($\phi$) threshold, 
and also in the region above the $c\bar{c}$ and $b\bar{b}$ narrow resonances.  

\index{duality}
Duality has also been applied successfully to describe the resonance region in
electron scattering data (Bodek and Yang 2002), and a comparison of a
quark model with actual electron scattering data is shown in
Figure~\ref{fig:bodek-yang}.  The resonance structure essentially
appears as modulations on top of the quark model prediction.  This
approach is now being used in most modern neutrino generators
attempting to interpolate the region of hadronic mass squared, $W^2$, 
best treated as
production of discrete resonances and the higher energy region where
parton model calculations of deep inelastic scattering are good
approximations.

\subsection{The Effect of the Nucleus}

\index{cross section!Fermi smearing}
A significant complication for the understanding of neutrino
interactions in future experiments is the need to model cross sections
on a variety of nuclei.  Exclusive neutrino interactions in the GeV
energy range of the type that must be understood in future experiments
are particularly sensitive to modification in the nuclear medium, and
there are few definitive models and little data currently available to help to
understand the effects.  In this
section, we will survey some of the relevant phenomenology.

One effect that is relevant for all energy regimes is the motion of
the target nucleon to the nucleus.  This is often called ``Fermi
smearing'', and it can have a dramatic effect.
Figure~\ref{fig:nuclear-smear} illustrates how significant this effect
is on the production of resonances in photo-nuclear absorption.  The
proton data represented by the solid line clearly shows multiple
resonance peaks, but in $^{12}C$, the same resonances become
indistinct due to Fermi smearing except for the well separated
$\Delta(1232)$ resonance.  Similar dramatic effects can be
seen in reconstruction of quasi-elastic events, and in scattering from
high $x$ partons in deep inelastic scattering.

\begin{figure}
\centering
\mbox{
\includegraphics[width=7cm]{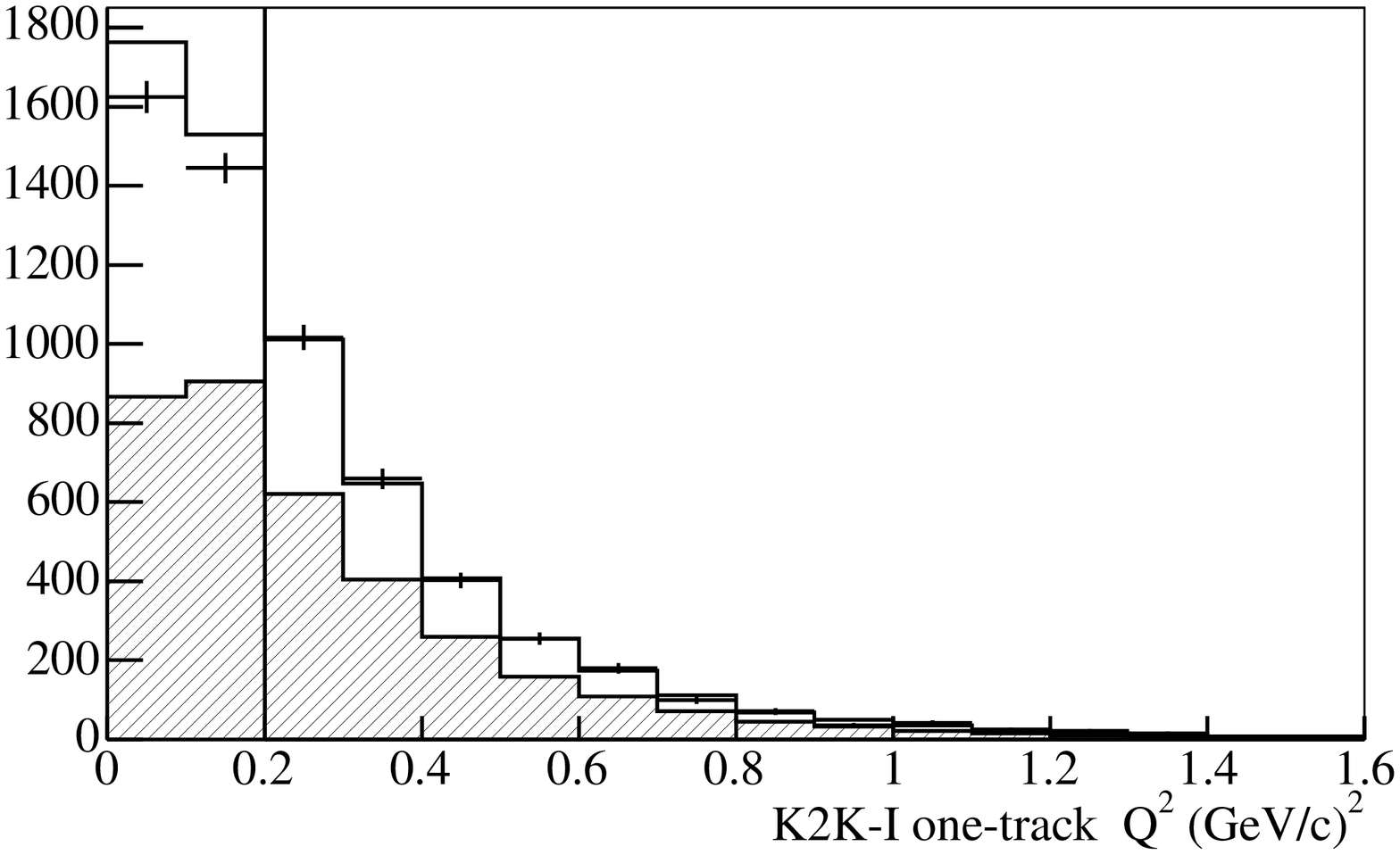}
\includegraphics[width=7cm]{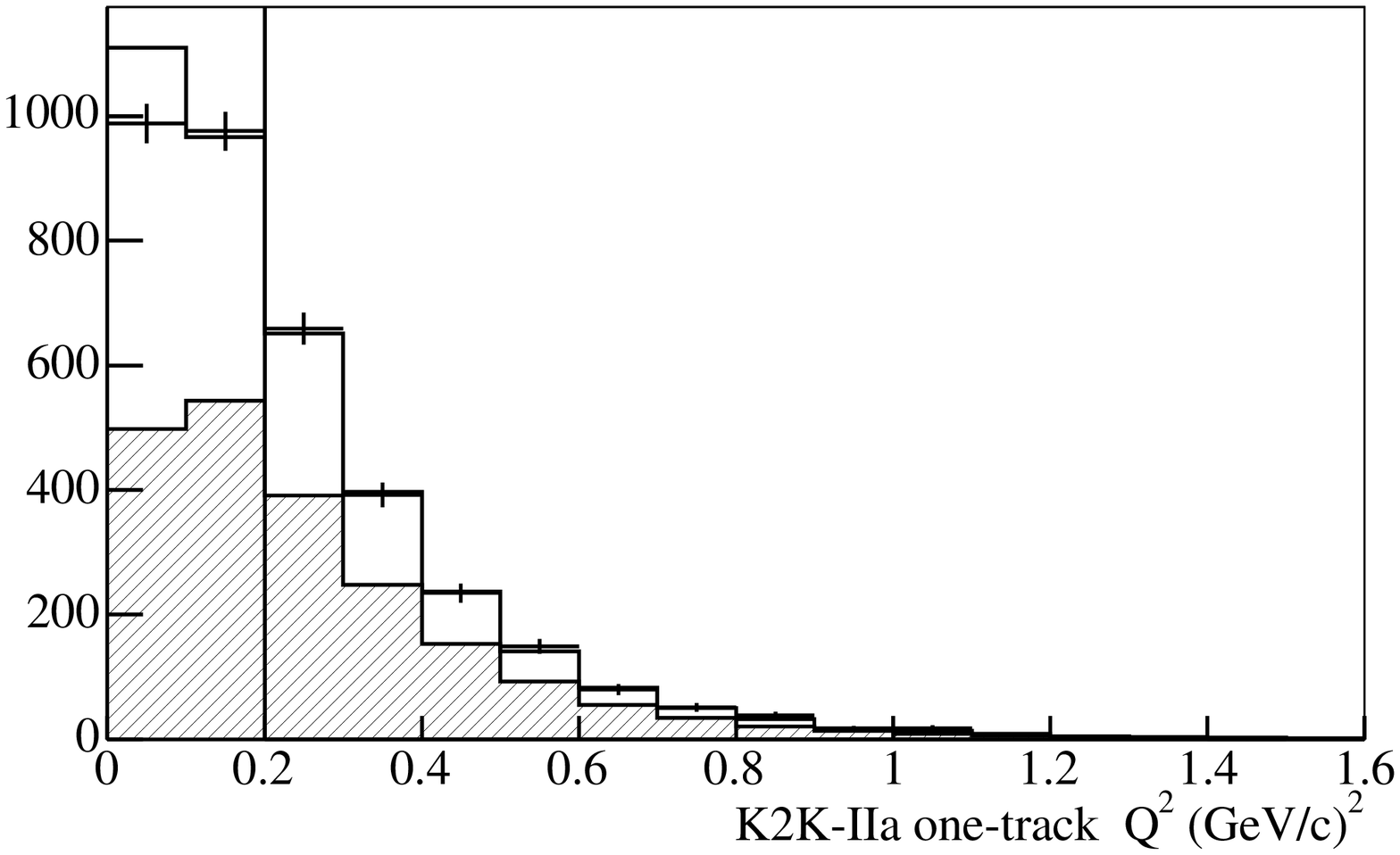}
}
\caption{$Q^2$ distribution in a quasi-elastic enriched sample in the SciBar
  detector at KEK (Gran 2006) shown for two different running periods.  
  The estimation of the number of events due to
  quasi-elastic processes in each bin is shown by the shaded area.}
\label{fig:scibar-QE}
\end{figure}

Quasi-elastic scattering at low $Q^2$ has a unique nuclear effect.
Because the final state nucleon will not necessarily be energetic
enough to leave the nucleus, its creation may be suppressed if there
is no free nuclear state available due to the Pauli exclusion
principle.  This effect is often called Pauli blocking.  The effect
can be modeled, although it is unclear how well these models work or
how universal they are.  Even with a model for Pauli blocking in
their predictions, both the SciBar and MiniBooNE detectors see
significant deficits of events from scattering off of carbon at low
$Q^2$ (Figure~\ref{fig:scibar-QE}).

Another significant effect is the rescattering in the nuclear medium
of hadrons which are sufficiently energetic to escape the target nucleus.
Because the nucleus is so dense, the material traversed when a produced
particle escapes the target nucleus is significant compared to the amount of
nuclear matter it sees when traveling through macroscopic amounts of
detector material afterward.  Therefore, it is not surprising that the
probability of a reinteraction in the nucleus is significant.  Such a
reinteraction may be particularly difficult for an experiment relying
on knowledge of exclusive final states, such use of two-body kinematic
constraints in reconstructing quasi-elastic events, or when concerned
about backgrounds to $\nu_e$ appearance from $\pi^0$s.  Again, these
effects are not well studied, although there is some promise in the
use of electron scattering data to constrain models of such final
state reinteractions.

\begin{figure}
\centering
\includegraphics[width=14cm]{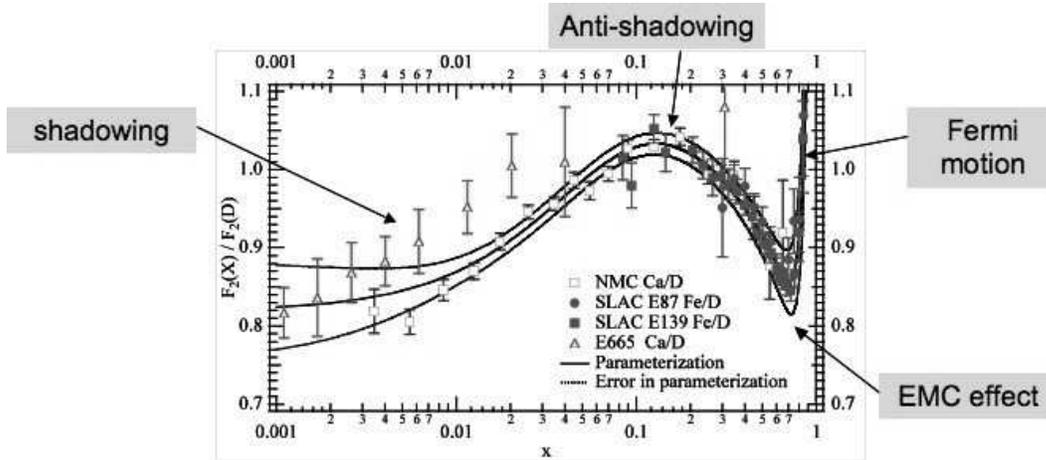}
\caption{Nuclear effects on parton distributions.}
\label{fig:EMC-effect}
\end{figure}

\index{deep inelastic scattering}
\index{partons}
In the deep inelastic scattering region, nuclear effects are well
measured in charged lepton scattering and are often parameterized in
terms of their effect on parton distribution functions as shown in
Figure~\ref{fig:EMC-effect}.  At high $x$, the same Fermi smearing
described above leads to a dramatic increase in the rare partons
carrying very high $x$.  The region of moderate $x$, in the ``valence
quark'' region is suppressed through an effect generally named the
``EMC'' effect after the first experiment to observe it.  At
$x\sim0.1$, there is a small enhancement of the PDFs sometimes
referred to as ``anti-shadowing'' and PDFs at low $x$ appear to be
dramatically suppressed due to ``shadowing''.  Because these effects
have only been measured in charged-lepton scattering and because, with
the exception of the Fermi smearing and perhaps shadowing, there are 
plausible but not definitive theoretical interpretations of
the effect, it is not clear whether the modifications to PDFs are in
fact universal, or whether the effects in neutrino neutral and
charged-current scattering will be different.  Most
likely, data from neutrino scattering experiments on a variety of
nuclei, including light nuclei, will be required to resolve this
question.

\subsection{Other Regimes of Transition}

There are other regions of transition, usually associated with binding
thresholds.  Binding energies of electrons in atoms are $\stackrel{<}{\sim}
Z^2m_ec^2\alpha_EM$ which can cover a broad range in energy from a few
eV to $10^5$~eV, and certainly at very low energies these bindings can
affect neutrino scattering from atomic electrons.  However, this is
not in an energy range that has effected neutrino oscillation
experiments to date.  There is also a
transition region associated with the binding energy of nucleons
inside the nucleus that ranges from $0.1$--$10$~MeV.  This binding
energy most definitely has had an impact on oscillation physics in a
number of experiments.  The SNO experiment uses charged and
neutral-current reactions $\nu_e d\to ppe^-$ and $\nu d\to pn\nu$ on
deuterons and elastic scattering from atomic electrons as its
oscillation signatures.  For the few MeV neutrinos from the sun, the
thresholds of atomic electrons, $<1$~keV even for oxygen, are
irrelevant.  However, the $2.2$~MeV binding energy of the deuteron and
the characteristic sharp quadratic turn-on with neutrino energy at the
threshold is a significant theoretical uncertainty in reaction rates
for low energy neutrinos.

\begin{figure}
\centering
\includegraphics[width=8cm]{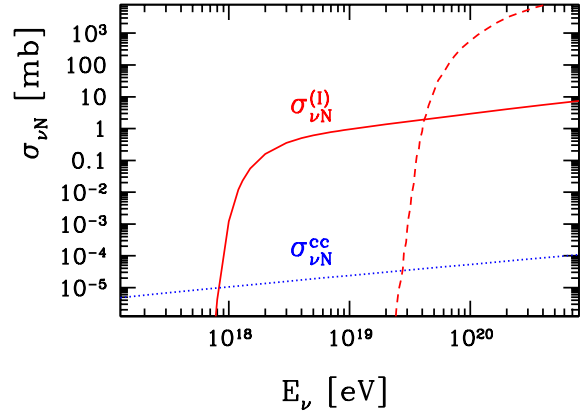}
\caption{Possible modifications of ultra-high neutrino energy
  cross sections could be quite dramatic if new degrees
  of freedom can be excited at high energies.  This model (Fodor 2003) shows an
  example using electroweak instantons.}
\label{fig:fodor}
\end{figure}

A more interesting possibility might be realized in cosmic ray physics
involving neutrinos.  Although our current knowledge from energy
frontier colliders limits the energy scale at which quarks and leptons
might be bound states to $\stackrel{>}{\sim}10$~TeV, beyond this it is
possible that some new process turns on as illustrated 
in Figure~\ref{fig:fodor}.  It is interesting to
remember from our discussions that the ``background'' deep inelastic
cross section will continue to grow with energy approximately linearly
until $Q^2>M_W^2$, when the propagator term (Equation~\ref{eqn:prop}) will
begin to drop with increasing $Q^2$.  This effect begins to be
noticeable at neutrino energies of $\sim10$~TeV, and is so significant
at high energies that baseline ``QCD'' cross section in
Figure~\ref{fig:fodor} is barely increasing with energy.

\section{Conclusions}

By way of conclusions, I will offer in compact form what I think are the most
important points for the student to take away from these lecture
notes.

The understanding of neutrino interactions is one of the keys to
precision measurements of neutrino oscillations at accelerators.  It
will soon limit precision in the current program of $\nu_\mu$
disappearance at atmospheric baselines. In the future
cross section uncertainties, if not addressed with new data, will play
a significant role in the ultimate precision of the $\nu_\mu\to\nu_e$
and $\nub_\mu\to\nub_e$ measurements needed to untangle the neutrino
mass hierarchy and to search for leptonic CP violation.

The neutrino scattering rate is generally proportional to energy.
Specifically this is true for scattering from pointlike particles when
the $Q^2\ll M_W^2$.  When it is not true for an exclusive process below
this $Q^2$ threshold, then there is some physics limiting the maximum
momentum transfer, such as a threshold above which the target breaks
up.

Neutrino target structure (atom, nucleus, nucleon) is a significant
complication to precise theoretical calculation of cross section on
neutrinos, particularly near inelastic thresholds.  Tools like
quark-hadron duality are helpful for modeling the major features, but
detailed predictive models require additional data we do not currently
have in hand, particularly when knowledge of the nuclear environment
is needed to make a firm prediction.

\section*{Acknowledgments}
I am grateful to the organizers of summer school for providing a
stimulating environment and program of study and for their hard work in
recruiting students to the school.   I thank in particular Franz
Muheim for his thoughtful and careful review and editing of this manuscript.
I also thank Dave Casper, Rik Gran, Debbie Harris, Jorge Morfin,
Tsuyoshi Nakaya and Sam Zeller for providing material for or useful
comments on the lectures on which this manuscript is based.

\section*{References}
\frenchspacing
\begin{small}
\reference{Bethe, H. and Peierls, R. (1934), {\it Nature} {\bf 133}, 532.}
\reference{Bodek, A. and Yang, U.-K. (2002), {\it Nucl.\ Phys.\ Proc.\
    Suppl.} {\bf 112} 70.}
\reference{Fermi, E. (1934), {\it Z. Physik} {\bf 88}, 161.}
\reference{Fodor, Z. {\em et al.} (2003), {\it
    Phys.\ Lett.} {\bf B561} 191.}
\reference{Galison, P. (1983), {\it Rev.\ Mod.\ Physics} {\bf 55}, 477.}
\reference{Gran, R. {\em et al.} (2006), {\it Phys.\ Rev.} {\bf D74}, 052002.}
\reference{Kretzer, S. and Reno, M.H. (2002), {\it Phys.\ Rev.}
   {\bf D66} 113007.}
\reference{Llewellyn Smith, C.H. (1972), {\it Phys.\ Rep.} {\bf 3C}, 261.}
\reference{McAlister, R. and Hofstadter, R. (1955), {\it Phys.\ Rev.}
    {\bf 102} 851.}
\reference{Minakata, H. and Nunokawa, H. (2001), {\it Jour.\
    HEP} {\bf 0110} 001.}
\reference{Reines, F. (1996), {\it Rev.\ Mod.\ Physics} {\bf 68} 317.}
\reference{Sterman, G. {\em et al.} (1995), {\it Rev.\ Mod.\ Physics} {\bf 67}, 157.}
\reference{Zeller, G.P. (2003), {\it arXiv:hep-ex}~0312061.}
\end{small}
\end{document}